\documentclass[aps,prx,twocolumn,english,showpacs,superscriptaddress,amssymb,amsfonts]{revtex4}
\usepackage{graphicx}
\usepackage{hyperref}
\newcommand{\Rom}[1]{\uppercase\expandafter{\romannumeral #1\relax}}

\begin{document}
	\title{Deciphering the nonlocal entanglement entropy of fracton topological orders}
	
	\author{Bowen Shi}\author{Yuan-Ming Lu}
	\affiliation{Department of Physics, The Ohio State University, Columbus, OH 43210, USA}
	
	\date{\today}
	
	\begin{abstract}
		The ground states of topological orders condense extended objects and support topological excitations. This nontrivial property leads to nonzero topological entanglement entropy $S_{topo}$ for conventional topological orders. Fracton topological order is an exotic class of models which is beyond the description of TQFT. With some assumptions about the condensates and the topological excitations, we derive a lower bound of the nonlocal entanglement entropy $S_{nonlocal}$ (a generalization of $S_{topo}$). The lower bound applies to Abelian stabilizer models including conventional topological orders as well as type \Rom{1} and type \Rom{2} fracton models, and it could be used to distinguish them. For fracton models, the lower bound shows that $S_{nonlocal}$ could obtain geometry-dependent values, and $S_{nonlocal}$ is extensive for certain choices of subsystems, including some choices which always give zero for TQFT. The stability of the lower bound under local perturbations is discussed.
	\end{abstract}

	\maketitle

\section{Introduction}
Topological order \cite{PhysRevB.41.9377} is a gapped quantum phase of matter beyond the description of the Landau-Ginzburg theory of symmetry breaking. Many of the early examples of topological orders (which we will refer to as conventional topological orders) share the following properties: robust ground state degeneracy which depends on the topology of the manifold \cite{PhysRevB.41.9377}, the ground states are locally indistinguishable \cite{PhysRevB.41.9377,bravyi2010topological,bravyi2011short}, the existence of integer dimensional condensates and logical operators that can be topologically deformed \cite{PhysRevB.71.045110}, nontrivial braiding statistics of anyons or other \emph{topological excitations} (or topologically charged excitations)  e.g., excitations which could not be created alone by local operators \cite{bravyi2011topological, Haah2013}, they are effectively described by topological quantum field theory (TQFT) at low temperatures \cite{PhysRevB.41.9377}, and they can be used to do fault-tolerant quantum information processing \cite{Kitaev20032}. And it is well known that, in 2D, a suitable linear combination of entanglement entropy with local contributions canceled is a topological invariance called the topological entanglement entropy  \cite{PhysRevLett.96.110404,PhysRevLett.96.110405}. Topological entanglement entropy is a property of the ground state wave function and it has been used to identify quantum spin liquid phases \cite{isakov2011topological}. It also contains information about the ground state degeneracy \cite{PhysRevLett.111.080503} and the forms of low-energy excitations \cite{2015PhRvB..92k5139K}. Generalizations of topological entanglement entropy into 3D bulk \cite{PhysRevB.84.195120} and boundary \cite{2015PhRvB..92k5139K} are studied.

On the other hand, there are recently discussed 3D exotic topological ordered models \cite{PhysRevLett.94.040402,bravyi2011topological,PhysRevA.83.042330,PhysRevB.88.125122,PhysRevB.92.235136,PhysRevB.94.235157,2016PhRvB..94o5128W,Pretko2017,Vijay2017,Ma2017,Hsieh2017,Slagle2017} that do not fit very well into the pictures above. These models have recently been classified into fracton topological orders \cite{PhysRevB.94.235157}.
While fracton models have locally indistinguishable ground states when placed on nontrivial manifolds and the ground state degeneracy is robust under local perturbations \cite{bravyi2010topological,bravyi2011short}, the ground state degeneracy depends on the system size (geometry) rather than merely the topology of the manifold \cite{bravyi2011topological,Haah2013,PhysRevB.88.125122}. While fracton models possess topological excitations \cite{bravyi2011topological, Haah2013}, these topological excitations are  constrained to move in lower dimensional submanifolds rather than the whole system \cite{PhysRevB.92.235136,PhysRevB.94.235157}. The condensates and logical operators can be fractal dimensional \cite{PhysRevB.88.125122,PhysRevB.94.235157} instead of integer dimensional. These models are beyond the description of TQFT.

There are type \Rom{1} and type \Rom{2} fracton topological orders. The type \Rom{1} fracton models include the Chamon-Bravyi-Leemhuis-Terhal (CBLT) model \cite{PhysRevLett.94.040402,bravyi2011topological}, the Majorana cubic model \cite{PhysRevB.92.235136}, and the X-cube model \cite{PhysRevB.94.235157}, etc.; they have integer dimensional condensates and logical operators. The type \Rom{2}  fracton models include Haah's code \cite{PhysRevA.83.042330} and many of the  fractal spin liquid models \cite{PhysRevB.88.125122} (see Sec.\ref{fractal model}). Type \Rom{2} fracton models possess fractal condensates and logical operators and the excitations are fully immobile \cite{PhysRevB.94.235157}.

For the ground state entanglement properties of fracton models, a relation to the ground state degeneracy is implied in \cite{PhysRevLett.111.080503} and the entanglement renormalization  group transformation of Haah's code is studied in \cite{haah2014bifurcation}. In this work, we construct a direct analogy of topological entanglement entropy by doing linear combinations of entanglement entropies of different subsystems in such a way that the local contributions (from each boundary or corner of the subsystems) are canceled, and we call the linear combination $S_{nonlocal}$, the \emph{nonlocal entanglement entropy}. While $S_{nonlocal}$ is topologically invariant for conventional topological orders, it is geometry-dependent for fracton models.

\begin{figure}[h]
	\centering
	\includegraphics[scale=0.350]{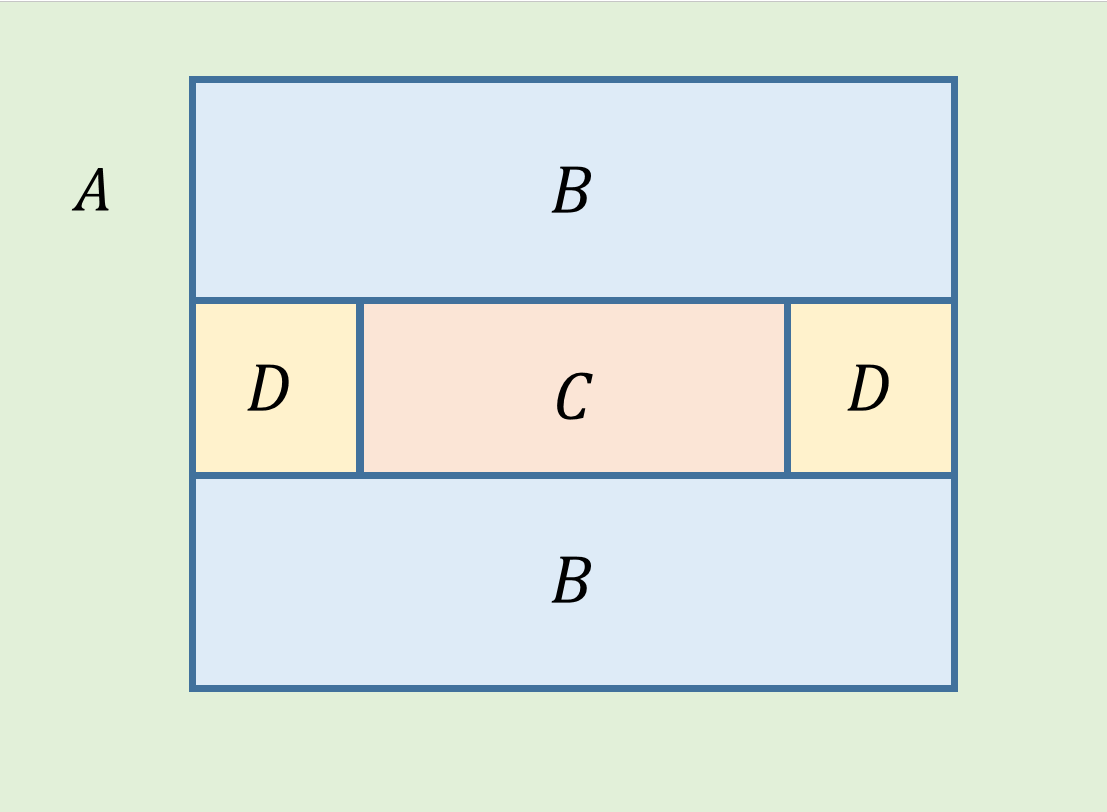}
	\caption{A system is divided into subsystems $A,B,C$ and $D$. Geometrically and topologically distinct choices will be used throughout the paper. They share the following features:
		$\partial A\cap\partial C=0$ and other pairs of subsystems have shared boundaries.}\label{ABCD}
\end{figure}

Explicitly, we choose a conditional mutual information form used by Kim and Brown \cite{2015PhRvB..92k5139K} and define the nonlocal entanglement entropy $S_{nonlocal}\equiv (S_{BC}+S_{CD}-S_C-S_D)\vert_{\rho}=I(A:C\vert B)\vert_{\rho}$. Where $\rho=\vert\psi\rangle\langle \psi\vert$ with $\vert \psi\rangle$ being a ground state.  The whole system is the union of subsystems $A$, $B$, $C$, and $D$ with  $A$ and $C$ separated by distance $l\gg \xi$, the correlation length. One can check that the local contributions from the boundaries are canceled, and this is why we use the name ``nonlocal entanglement entropy." This construction can be used in any dimensions and an example in 2D is shown in Fig.\ref{ABCD}. With several assumptions about the condensates and topological excitations, a lower bound of $S_{nonlocal}$ is derived.

When applied to known conventional Abelian topological orders e.g. the 2D toric code model and the 3D toric code model \cite{2002JMP....43.4452D}, the lower bound is topologically invariant and it is identical to the exact result, i.e. the lower bound is saturated. When applied to fracton models \cite{PhysRevLett.94.040402,bravyi2011topological,PhysRevA.83.042330,PhysRevB.88.125122,PhysRevB.92.235136,PhysRevB.94.235157},  the lower bound depends on the sizes and relative locations of the subsystems. For fracton models, there exist choices of subsystems for which the lower bound is nonzero and extensive. It is possible to have $S_{nonlocal}>0$  for subsystem choices, that are expected to have $S_{nonlocal}=0$ if TQFT holds.

This method observes an intimate relation between $S_{nonlocal}$ and topological excitations created in $D$ by a unitary operator $U$ stretched out in $CD$ which could be ``deformed" into a unitary operator $U^{def}$ in $AD$, and since deformable $U$ is intimately related to condensate operator $W$ (for more details of $U$, $U^{def}$ and condensate operator $W$ see Sec.\ref{The main result.}), this method observes an intimate relation between $S_{nonlocal}$ and ground state condensates as well.
 This method allows a lower bound of $S_{nonlocal}$ to be obtained without calculating the entanglement entropy of any individual subsystem.
Furthermore, it provides us with a unified viewpoint to understand the topology-dependent $S_{nonlocal}$ in the conventional topological orders and the geometry-dependent $S_{nonlocal}$ in fracton topological orders. Also discussed is the stability of the lower bound under local perturbations.

Two additional papers appeared after our work which also study the entanglement entropy of fracton phases, using explicit computation  \cite{2017arXiv171001744M} and tensor network \cite{2017arXiv171004220H}.

For non-Abelian models, some of our assumptions breakdown, and  our original method does not apply. Nevertheless, a variant of our lower bound is applicable to non-Abelian models \cite{2018arXiv180101519S}.

The structure of the paper is as follows: In Sec.\ref{Sec.1} we provide a derivation of the lower bound from some assumptions about topological excitations and condensates; In Sec.\ref{Sec.2} we apply our lower bound to several exactly solved Abelian stabilizer models of 2D, 3D conventional topological orders and type \Rom{1}, type \Rom{2} fracton topological orders. In Sec.\ref{Sec.3} we discuss the stability of the lower bound under local perturbations. Sec.\ref{Sec.4} is discussion and outlook.

\section{The lower bound}\label{Sec.1}
\subsection{A few notations and definitions}
We will consider a infinite system without boundaries. The system is divided into subsystems $A$, $B$, $C$, $D$ (nonoverlapping regions in real space, the union of which is the whole system). Each subsystem has a size large compared to the correlation length $\xi$, and the subsystems $A$ and $C$ are separated by a distance much larger than the correlation length. One example is shown in Fig.\ref{ABCD}, and similar constructions can apply to any dimensions. For all the examples in this paper, we have chosen $B$, $C$, $D$ to be local subsystems while $A$ is not, but there exist other possible choices, say $A$, $B$, $C$ local. A local subsystem is a subsystem which can be contained in a ball-shaped subsystem of finite radius. $\partial A$, $\partial B$ are the boundaries of the subsystems. We use $\bar{A}$ to denote the complement of $A$.

We will use $\rho$, $\sigma$ to represent density matrices. In this paper we always use $\rho$ for the ground state density matrix, $\rho=\vert\psi\rangle\langle \psi\vert$, and $\vert\psi\rangle$ is the ground state. We use $\rho_{ABC}$, $\sigma_{BC}$  when we want to specify the subsystems. The entanglement entropy is defined in terms of the (reduced) density matrix as usual
$S=-\textrm{tr} [\sigma\ln \sigma]$.
 We  use  $S_{ABC}\vert_\rho$ and $S_{ABC}\vert_\sigma$ to distinguish the entanglement entropy on region $ABC$ with different density matrices $\rho$, $\sigma$.
Define  conditional mutual information
\[
I(A:C\vert B)\equiv S_{AB}+S_{BC}-S_B-S_{ABC}
\]
and we use $I(A:C\vert B)\vert_\rho$ when we want to specify a density matrix. It is known that the conditional mutual information is always nonnegative $I(A:C\vert B)\ge 0$. We say $\sigma_{ABC}$ is conditionally independent if $I(A:C\vert B)\vert_\sigma=0$.

For unitary operators  $U$ and $U'$ which create excitations in $D$ when acting on the ground state $\vert\psi\rangle$, we say $U\sim U'$ or $U$ is similar to $U'$ if the states $U\vert \psi\rangle$ and $U'\vert \psi\rangle$ have identical reduced density matrices on $ABC$, i.e. $tr_D [U\rho U^\dagger]= tr_D [U'\rho U'^\dagger]$. Otherwise, we say $U$ and $U'$ are \emph{distinct}.

\subsection{Prepare for the lower bound}
If there is a density matrix $\sigma$ which is related to the ground state density matrix $\rho$ by $\rho_{AB}=\sigma_{AB}$ and $\rho_{BC}=\sigma_{BC}$, then we have:
\begin{equation}
I(A:C\vert B)\vert_\rho\ge S_{ABC}\vert_\sigma-S_{ABC}\vert_\rho \label{the key}
\end{equation}
and the ``$=$" happens if and only if $I(A:C\vert B)\vert_\sigma= 0$. For a proof, observe that $I(A:C\vert B)\vert_\rho$ and $I(A:C\vert B)\vert_\sigma$ has only a single different term, and that $I(A:C\vert B)\vert_\sigma\ge 0$.

For a pure state, the entanglement entropy of a subsystem equals the entanglement entropy of its complement, e.g. $S_{\Omega}=S_{\bar{\Omega}}$ for any subsystem $\Omega$. Therefore:
\[
I(A:C\vert B)\vert_{\rho}= (S_{BC}+S_{CD}-S_B-S_D )\vert_{\rho}.
\]

Observe that the local contributions of the entanglement entropy get canceled due to the fact that $A$ and $C$ are separated. Let us define the nonlocal entanglement entropy (of the ground state)
\begin{equation}
S_{nonlocal}\equiv (S_{BC}+S_{CD}-S_B-S_D)\vert_{\rho}. \label{the local form}
\end{equation}
The nonlocal entanglement entropy is just another way to write down the conditional mutual information, $S_{nonlocal}=I(A:C\vert B)\vert_\rho$, and therefore $S_{nonlocal}\ge 0$.
The form in Eq.(\ref{the local form}) has the advantage that it involves only local systems $B$, $C$, $D$. When the system is placed on a torus or other nontrivial manifolds instead of a infinite manifold, the system may have several locally indistinguishable ground states, this form of $S_{nonlocal}$ in terms of local subsystems is more convenient, and even if $\rho$ is a mixed state density matrix of different locally indistinguishable ground states, $S_{nonlocal}$ still has the same value.

\subsection{The key idea about the lower bound}\label{the key idea}
The discussion above suggests a way to obtain a lower bound of $S_{nonlocal}$. For any $\sigma$ satisfying  $\sigma_{AB}=\rho_{AB}$ and $\sigma_{BC}=\rho_{BC}$,
\begin{equation}
S_{nonlocal} \ge S_{ABC}\vert_\sigma-S_{ABC}\vert_\rho.     \label{the lower bound}
\end{equation}
The density matrix $\sigma$ does not have to be a density matrix of a pure state.
If we could find a $\sigma$ satisfying the above requirement and  $S_{ABC}\vert_\sigma>S_{ABC}\vert_\rho$,  a nonzero lower bound is obtained, and then the existence of nonzero nonlocal entanglement entropy is established.

Now, let us assume that we could find a set of $\sigma_I$ with $I=1,\ldots,N$ such that $\sigma_{I\,AB}=\rho_{AB}$ and $\sigma_{I\,BC}=\rho_{BC}$. Then we can do superpositions and define ${\sigma}\equiv \sum_{I=1}^N p_I \sigma_I$ with $\{p_I \}$ being a probability distribution, i.e. $p_I\in[0,1]$ and $\sum_{I=1}^N p_I=1$. The ${\sigma}$ is a new density matrix which satisfies ${\sigma}_{AB}=\rho_{AB}$ and ${\sigma}_{BC}=\rho_{BC}$. So we have a whole parameter space of ${\sigma}$ to try.

If lucky, we may even be able to find a $\sigma^\ast_{ABC}$ which is conditionally independent (satisfying $I(A:C\vert B)\vert_{\sigma^\ast}=0$) and  we  have an exact result
\begin{equation}
S_{nonlocal}= S_{ABC}\vert_{\sigma^\ast}-S_{ABC}\vert_\rho.
\end{equation}
Or, if we find the lower bound is saturated, we know the $\sigma$ we used to obtain the lower bound is conditionally independent.
We note that, in the quantum case (unlike the classical case), it is not always possible to find a conditionally independent $\sigma^\ast$ such that $\sigma^\ast_{AB}=\rho_{AB}$ and $\sigma^\ast_{BC}=\rho_{BC}$ for $\rho$ being a general density matrix \cite{Ibinson2008}. 
Therefore, the existence of such $\sigma^\ast$ in some system might be interesting by itself. On the other hand, the conditional independent state $\sigma^\ast$ is known to exist for models satisfying simple conditions  (\Rom{1})(\Rom{2}) in \cite{2016PhRvA..93b2317K}. 

\subsection{Calculate the lower bound for Abelian models employing assumptions}\label{The main result.}
We make a few  assumptions about the condensates and operators creating topological excitations
in order to develop a  way to find  $\sigma_I$ and calculate a lower bound of $S_{nonlocal}$. These assumptions are applicable to Abelian models with commuting projector Hamiltonians (with each term acting on a few sites localized in real space), including conventional models e.g. the toric code model in various dimensions \cite{2002JMP....43.4452D}, quantum double models with any Abelian finite group, and  fracton models \cite{PhysRevLett.94.040402,bravyi2011topological,PhysRevA.83.042330,PhysRevB.88.125122,PhysRevB.92.235136,PhysRevB.94.235157} of type \Rom{1} and type \Rom{2}  which we will be focusing on in this work. In the context of fracton models, Abelian means no protected degeneracy associated with excitations.

Our way of employing the operators is inspired by a method by Kim and Brown \cite{2015PhRvB..92k5139K}, where an interesting connection between conditional mutual information and deformable operator $U$ is obtained.
While the subsystems we choose have only an unimportant difference from \cite{2015PhRvB..92k5139K}, our result is different. The result in \cite{2015PhRvB..92k5139K} shows that if $S_{nonlocal}=0$, there will be no topological excitations, and therefore, $S_{nonlocal}>0$ is needed for the existence of topological excitations. The result is very general since very small amount of assumptions was used. Nevertheless, the result was not powerful as a lower bound for $S_{nonlocal}$. In this work, on the other hand, we use more detailed properties of topological excitations and condensates to obtain a powerful lower bound of $S_{nonlocal}$. Our method shows that the key to have $S_{nonlocal}>0$ in these models is the nonlocal nature of the ground state condensates and the operators creating topological excitations. $S_{nonlocal}$  can be extensive in the subsystem size and it is not necessarily  topologically invariant. Whether $S_{nonlocal}$ is topologically invariant or not depends on whether the operators can be deformed topologically.

\begin{figure}[h]
	\centering
	\includegraphics[scale=0.310]{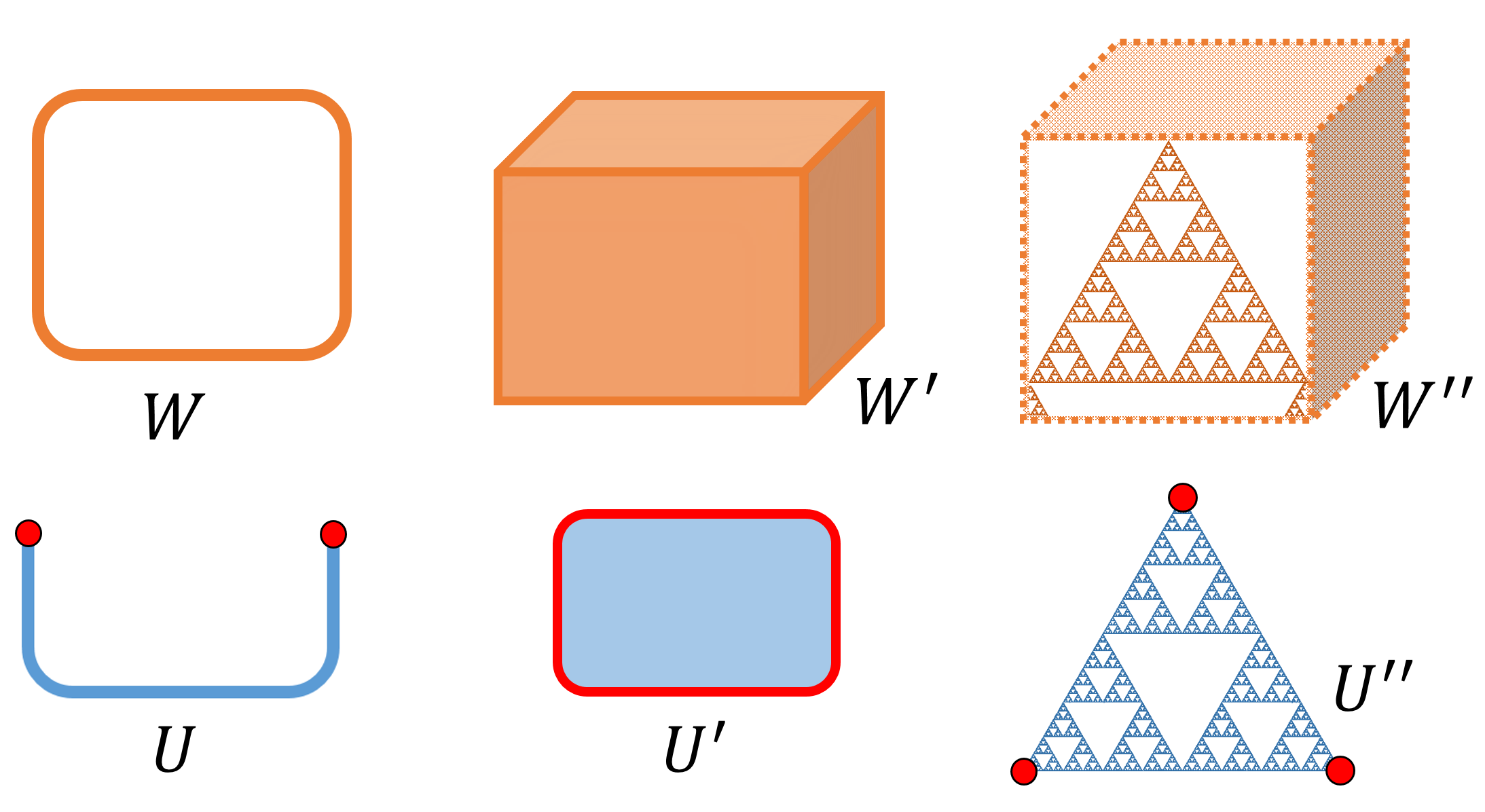}
	\caption{Condensate operators from different topological orders $W$, $W'$, $W''$ (in orange color), and the truncations of  corresponding operators give us deformable operators $U$, $U'$, $U''$ (in blue color) which create topological excitations (in red color) when acting on the ground state. The operators $W$, $U$ are from the 2D toric code model; the operators $W'$, $U'$ are from the 3D toric code model; the operators $W''$, $U''$ are from a fractal spin liquid model (note that $W''$ is not a precise depiction). The support of $W$ is a closed loop; the support of $W'$ is a closed membrane, i.e. the 2D boundary of the box; the support of $W''$ has fractal structure. The picture only shows part of $W''$ explicitly, the rest of $W''$ is embedded in the dotted 2D surfaces with dashed 1D edges, and it may contain fractal parts or 1D parts.}\label{Truncation}
\end{figure}

Before rigorously stating the assumptions and deriving the lower bound, here are a few words about the physical picture.
The ground states of topological orders condense extended objects. If a unitary operator $W$ (which has an extended support) acting on the ground state $\vert\psi\rangle$ gives you $W\vert\psi\rangle =e^{i\varphi}\vert\psi\rangle$,  we call the operator $W$   a \emph{condensate operator} with eigenvalue $e^{i\varphi}$. Whenever confusion can be avoided, we may call $W$ a \emph{condensate} for short. Let us further assume $W$ to be a tensor product of operators acting on each site.
Then, a suitably defined ``truncation" of a condensate operator $W$ onto a subsystem $\Omega$ gives you a new operator $U$. $U\vert\psi\rangle$ is an excited state with  topologically excitations located around $\partial \Omega\cap W$, and $U$ can be deformed in the sense that you could choose a ``truncation" of $W^\dagger$ onto $\bar{\Omega}$ and call it $U^{def}$, which creates the same topological excitations and satisfies $U\vert\psi\rangle= U^{def}\vert\psi\rangle$. One the other hand, if we have unitary operators $U$ and $U^{def}$ satisfying $U\vert\psi\rangle= U^{def}\vert\psi\rangle$, then $[U^{def}]^\dagger U\vert\psi\rangle =\vert\psi\rangle$, and therefore $[U^{def}]^\dagger U$ is a condensate operator with eigenvalue $+1$. Intuitively, a condensate operator $W$ can be ``truncated" into a  deformable operator $U$ which creates topological excitations, and a pair of deformable operators $U$ and $U^{def}$ can be ``glued together" into a condensate operator $W$. Therefore $U$ and $W$ are closely related.
Some examples of condensate operator $W$ and deformable operator $U$,  are shown in Fig.\ref{Truncation}.
Once we have condensate operators $W_i$ and deformable operators $U_j$, we use $U_j$ to create topological excitations in $D$ which result in some states $\sigma_I$ which is identical to the ground state $\rho$ on $AB$ and $BC$. If the excitations created in $D$ are topological excitations, they can not be created by an operator supported on $D$, and  $\sigma_{I\,ABC}$ and $\rho_{ABC}$ will have some difference. The difference is detected by a change of the eigenvalue of condensate operator $W_i$ supported on $ABC$.  Then, we use $\sigma_I$ to obtain a lower bound of $S_{nonlocal}$.

\begin{figure}[h]
	\centering
	\includegraphics[scale=0.54]{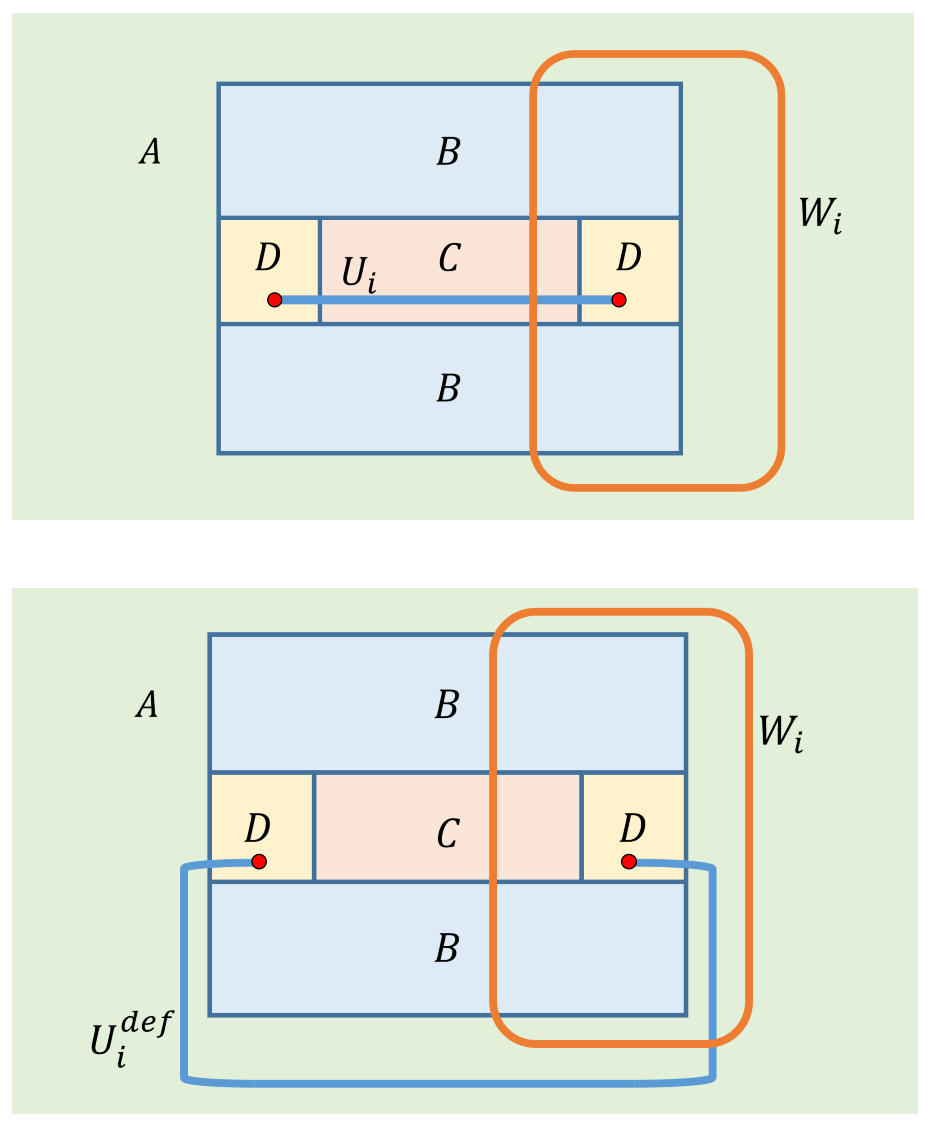}
	\caption{An illustration of the support of different operators: $W_i$ (in orange color) is supported on $ABC$; $U_i$ (in blue color) is supported on $CD$; $U_i^{def}$ (in blue color) is supported on $AD$. The topological excitations created by $U_i$ or $U_i^{def}$ are shown in red color. The color setting of the operators and excitations will be used throughout the paper.
		 We do not assume these operators to be integer dimensional and the construction applies to models in different dimensions.}\label{deform}
\end{figure}

The following are our assumptions $\mathbf{U}$-1, $\mathbf{U}$-2, $\mathbf{U}$-3; $\mathbf{W}$-1, $\mathbf{W}$-2:

\emph{Assumption} $\mathbf{U}$-1: There exists a set of unitary operators $\{U_i\}$  supported on $CD$ and a set of unitary operators  $\{ {U}_{i}^{def} \}$  supported on $AD$, with $i=1,\cdots, M$. See Fig.\ref{deform} for an example.
 When acting on the ground state $\vert \psi\rangle$, $U_i$ can be ``deformed" into $U_{i}^{def}$, i.e.:
\begin{equation}
U_i\vert\psi\rangle =U_{i}^{def}\vert\psi\rangle. \label{algebra1}
\end{equation}

\emph{Assumption} $\mathbf{U}$-2: For any subsystem $\Omega$, the unitary operator $U_i$ can always be written as a direct product of unitary operators $U_{i\, \Omega}$ and $U_{i\,\bar{\Omega}}$ which act on the subsystem $\Omega$, $\bar{\Omega}$ respectively:
\begin{equation}
U_i=U_{i\,\Omega}\otimes U_{i\,\bar{\Omega}}. \label{algebra p}
\end{equation}

\emph{Assumption} $\mathbf{U}$-3: There are integers $n_i$ such that $U_i^{n_i}=1$, and when multiple $U_i$ act on a ground state $\vert\psi\rangle$, we have the following:
\begin{equation}
\bigg(\prod_{i=1}^M U_i^{k_i} \bigg)\vert \psi\rangle =e^{i\delta(\{k_i\})} \bigg(\prod_{i=1}^M [U_{i}^{def}]^{k_i}\bigg)\vert\psi\rangle,  \label{algebra2}
\end{equation}
where integer $0\le k_i\le n_i-1$, and we allow possible phase factors $e^{i\delta(\{k_i \})}$.

\emph{Assumption} $\mathbf{W}$-1: There exists a set of unitary operators \{$W_i$\} supported on subsystem $ABC$ such that
\begin{equation}
W_i \vert\psi\rangle= \vert\psi\rangle\qquad\qquad i=1,\ldots,M. \label{algebra3}
\end{equation}

\emph{Assumption} $\mathbf{W}$-2: The following relation between $W_i$ and $U_j$ holds:
\begin{equation}
W_iU_j=U_jW_i e^{i\theta_{ij}}\qquad \textrm{with}\qquad \theta_{ij}=\frac{2\pi}{n_i}\delta_{ij},   \label{algebra4}
\end{equation}
where $\delta_{ij}$ is the Kronecker delta.

Comments about the assumptions:

1) $\mathbf{U}$-1 implies that when $U_i$ is acting on the ground state, it can create excitations only in $D$, but not in $ABC$.

2) We do not assume $U_i$, $W_i$ to be string operators and not even assume $U_i$, $W_i$ to be integer dimensional operators. In fact, we will apply this method to fractal operators later.

3) In $\mathbf{U}$-1 we assumed $U$ can be deformed without changing the excitations. Nevertheless, we do not assume $U$ can be deformed into all topologically equivalent configurations, and we do not assume $U$ can be deformed continuously. As we will see below, in fracton models,  deformations exist in weird form, $U$ may not be topologically deformed (i.e. deformed continuously into any topologically equivalent configuration), and $U$ can sometimes be deformed in discontinuous ways into topologically inequivalent configurations.

4) $\mathbf{U}$-2 is not true when a local perturbation is added. We will address the stability of the lower bound under local perturbations separately in Sec.\ref{Sec.3}.

5) For the non-Abelian case, the entanglement entropy of an excited state could depend on the quantum dimension of the anyon \cite{PhysRevLett.96.110404,2018arXiv180101519S}, and  $\mathbf{U}$-2 does not apply. On the other hand, the idea in Sec.\ref{the key idea}  still holds, a saturated lower bound for non-Abelian models is recently discussed in \cite{2018arXiv180101519S}.

6) For systems with boundaries, one may choose $D$ being region attached to boundaries, as is done in \cite{2015PhRvB..92k5139K}. An alternative way is to identify $D$ with  a boundary region $\partial \Omega$; in this case, $\mathbf{U}$-1 should be understood as: $U_i$ being an operator supported on $C$ and attached to $\partial C\cap\partial \Omega$, and $U_{i}^{def}$ being an operator supported on $A$ and attached to $\partial A\cap\partial\Omega$.

7) According to $\mathbf{W}$-2, $\vert\psi\rangle$ and $U_i\vert\psi\rangle$ are eigenstates of $W_i$ with different eigenvalues, where $\vert\psi\rangle$ is the ground state. Since $W_i$ is supported on $ABC$, this implies that $U_i$  is distinct from the identity operator. Similarly, $U_i$ and $U_j$ are distinct for $i\ne j$. We will refer to this change of eigenvalue of $W_i$ as a \emph{detection}, e.g. $U_i$ is detected by $W_i$. The requirement $\theta_{ij}=\frac{2\pi}{n_i}\delta_{ij}$ is not crucial, and it can be replaced by other numbers as long as  the operator set $\{ W_i \}$ can detect the difference among the set of operators $\{U_i \}$.

8) For a relatively simple class of models, which has the Hamiltonian $H=-\sum_k h_k$, $[h_i,h_j]=0$ and $h_k^2=1$, there is  an obvious class of operators that satisfy Eq.(\ref{algebra3}) in $\mathbf{W}$-1, namely $W_i=\prod_{k\in \mathcal{E}_i} h_k$, where $\mathcal{E}_i$ is a subset of the stabilizer generators. For the ground state $\vert \psi\rangle$, we have $h_i\vert \psi\rangle =\vert\psi\rangle$, it follows that $W_i\vert\psi\rangle=\vert\psi\rangle$. As $U_i$ could not flip stabilizers in $ABC$, $\mathcal{E}_i$ must contain some $h_j$ in $D$. It turns out that this simple observation applies to all the Abelian stabilizer models we will use as examples in Sec.\ref{Sec.2}. However, we do not provide a general procedure to find the subset $\mathcal{E}_i$ for fracton models.
On the other hand, our method  works  for models not in this simple class also, such as quantum double models \cite{Kitaev20032,bombin2008family} with Abelian finite groups.

Define the set of states
\[
\vert \{k_i\};\psi\rangle\equiv \prod_{i=1}^{M}U_i^{k_i}\vert\psi\rangle \qquad\textrm{with}\qquad 0\le k_i \le n_i-1
\]
with $k_i$ being integers. Define $\sigma(\{k_i\})\equiv \vert \{k_i\};\psi\rangle \langle\{k_i\};\psi\vert $.  Note the total number of $\sigma(\{ k_i\})$ is $\prod_{i=1}^M n_i$. Relabel $\sigma(\{ k_i\})$ using a new index $I=1,\cdots, N$ with $N=\prod_{i=1}^M n_i$ and call them $\sigma_I$. One immediately varifies that \\
1) $\mathbf{U}$-1, $\mathbf{U}$-3 $\Rightarrow$  $\sigma_{I\,AB}=\rho_{AB} $ and $\sigma_{I\,BC}=\rho_{BC}$;\\
2) $\mathbf{W}$-1, $\mathbf{W}$-2 $\Rightarrow$ $\sigma_{I\, ABC}\cdot\sigma_{J\,ABC}=0$ for $I\ne J$;\\
3) $\mathbf{U}$-2 $\Rightarrow$ $\sigma_{I\,ABC}=V_I \rho_{ABC}V_I^\dagger$ and $S_{ABC}\vert_{\sigma_I}=S_{ABC}\vert_{\rho}$.

Where $V_I$ is some unitary operator acting on subsystem $ABC$ and recall that $\rho$ is the ground state density matrix.

Let ${\sigma}=\sum_{I=1}^N p_I \sigma_I$ with probability distribution $\{p_I \}$,  one derives that
\[
S_{ABC}\vert_{{\sigma}}-S_{ABC}\vert_\rho=-\sum_{I=1}^N p_I\ln p_I \le \ln N \nonumber
\]
``=" if and only if $p_I=\frac{1}{N}$ for all $I$. From Eq.(\ref{the lower bound}) we find
\begin{equation}
S_{nonlocal}\vert_\rho \ge \ln N =\sum_{i=1}^M \ln n_i. \label{the N bound}
\end{equation}

\subsection{When is the lower bound  topologically invariant?}\label{topological invariant}
It is instructive to think of the conditions under which our lower bound of $S_{nonlocal}$ is topologically invariant.

 Consider a chosen set of subsystems $A$, $B$, $C$, $D$ and the operator sets $\{U_i \}$, $\{U_{i}^{def} \}$ and $\{W_i\}$.
Let us do ``topological deformations" of the subsystems and  the operator sets.
Here, by ``topological deformation" of the operator  sets we mean that we can topologically deform the support of each operator to get new operator sets which preserve the algebra in Eq.(\ref{algebra1},\ref{algebra p},\ref{algebra2},\ref{algebra3},\ref{algebra4}). Note that, these deformations generally change the positions of the excitations, which should be contrasted with the type of deformation in $\mathbf{U}$-1, in which the positions of the excitations never change. When these conditions are satisfied, the lower bound for the two topologically equivalent choices of subsystems are the same. If such conditions are satisfied for each pair of topologically equivalent choices of subsystems, then our lower bound will be topologically invariant.

As is shown in the examples below in Sec.\ref{Sec.2}, $S_{nonlocal}$ can be either topologically invariant or not, and it is instructive to think of how the conditions above are violated in fracton models \cite{PhysRevLett.94.040402,bravyi2011topological,PhysRevA.83.042330,PhysRevB.88.125122,PhysRevB.92.235136,PhysRevB.94.235157} in which $S_{nonlocal}$ depends on the geometry of subsystems.

\section{Applications}\label{Sec.2}
In this section, our lower bound is applied to several stabilizer models of Abelian phases: the 2D and 3D conventional topological orders  and type \Rom{1}, type \Rom{2} fracton phases.

\subsection{The 2D Toric Code Model}
For a 2D topological order, choose the subsystems $A$, $B$, $C$, $D$ of the same topology as is shown in Fig.\ref{ABCD}. From the well-known results \cite{PhysRevLett.96.110404,PhysRevLett.96.110405}, one derives $S_{nonlocal}=2\gamma$ where $-\gamma$ is the topological entanglement entropy.
\begin{figure}[h]
	\centering
	\includegraphics[scale=0.30]{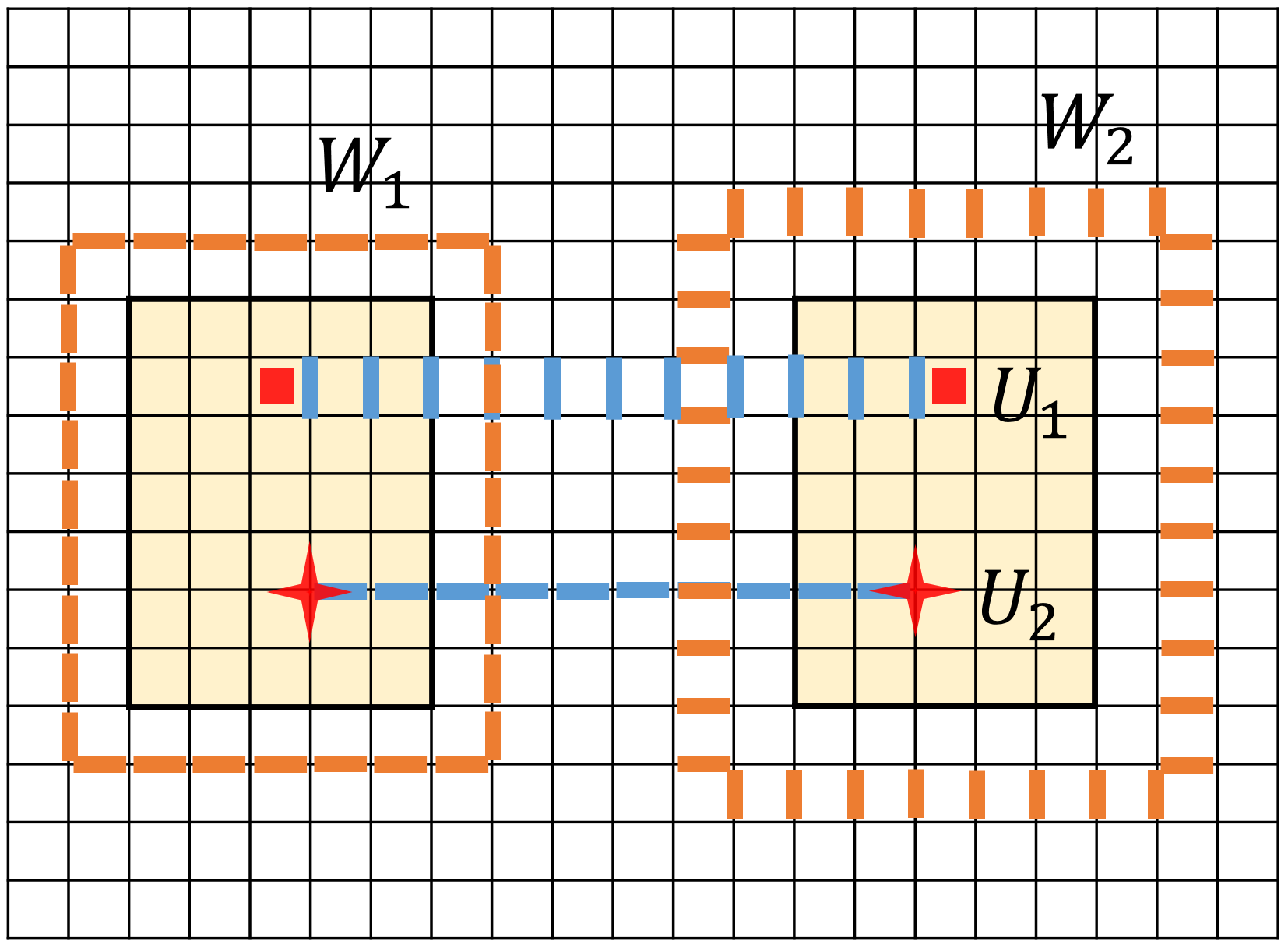}
	\caption{Deformable operators $U_1$, $U_2$ supported on open strings and condensate operators $W_1$, $W_2$ supported on closed strings.  $U_1$, $U_2$ create topological excitations around their endpoints. In other words, $U_1$ flips two plaquettes and $U_2$ flips two stars.  }\label{toric code}
\end{figure}

For the 2D toric code model \cite{Kitaev20032}: On a square lattice with a qubit on each link,  the Hamiltonian is
\[
H=-\sum_s A_s -\sum_{p} B_p,
\]
where $A_s$ is a product of $X_r$ of a ``star" or vertex, $B_p$ is a product of $Z_r$ of a plaquette.
\[
 A_s=\prod_{r\in s}X_r \qquad  B_{p}=\prod_{r\in p} Z_r,
\]
where $X_r$, $Z_r$ are Pauli operators acting on the qubit on link $r$.

The ground state of toric code model condenses two types of closed string operators, and the corresponding open string operators (which could be regarded as truncations of closed string operators) create topological excitations at the endpoints.

We find the following unitary operators $U_1$, $U_2$, $W_1$, $W_2$ as is shown in Fig.\ref{toric code}. $U_1$ and $W_2$ are products of $X_r$; $U_2$ and $W_1$ are products of $Z_r$. Also notice the feature that the closed string operators $W_1$  ($W_2$) can be written as a product of stabilizers $B_p$ ($A_s$) on a 2D disk region surrounded by the corresponding closed strings.

 $\mathbf{U}$-1, $\mathbf{U}$-2, $\mathbf{U}$-3, $\mathbf{W}$-1, $\mathbf{W}$-2 can be checked.
 The operators satisfy:
\begin{eqnarray}
&&U_1^2=U_2^2=W_1^2=W_2^2=1\nonumber\\
&&W_iU_j=U_jW_i e^{i\pi\delta_{ij}}\qquad\qquad i,j=1,2.\nonumber
\end{eqnarray}
Therefore $M=2$, $n_1=n_2=2$ and  $N=n_1n_2=4$. Using the result in Eq.(\ref{the N bound}), one derives $S_{nonlocal}\ge 2\ln 2$. By comparing with the known  result $\gamma=\ln 2$, $S_{nonlocal}=2\gamma=2\ln 2$, we find that our lower bound is saturated.

A by-product of a saturated lower bound is an explicit construction of a conditionally independent $\sigma^\ast$. In the toric code case:
\begin{equation}
\sigma^\ast =\frac{1}{4}(\rho +U_1 \rho U_1^\dagger + U_2 \rho U_2^\dagger + U_1 U_2 \rho U_2^\dagger U_1^\dagger)
\end{equation}
and $\sigma^\ast$ satisfies:\\
1) $\sigma^\ast_{AB}=\rho_{AB}$, $\sigma^\ast_{BC}=\rho_{BC}$; \\
2) $I(A:C\vert B)\vert_{\sigma^\ast}=0$.\\
Where $\rho=\vert\psi\rangle\langle \psi\vert$ is the ground state density matrix.

The observation in Sec.\ref{topological invariant} explains why the lower bound is topologically invariant in the toric code model: the operators $U_i$ and $W_i$ can be topologically deformed together with the subsystems $A$, $B$, $C$ and $D$, without changing the algebra in Eq.(\ref{algebra1},\ref{algebra p},\ref{algebra2},\ref{algebra3},\ref{algebra4}).

This method can be applied to other 2D Abelian topological orders, e.g., quantum double models with Abelian finite groups, and the lower bounds are saturated. For a variant of the method for non-Abelian models, see \cite{2018arXiv180101519S}.

\subsection{The 3D Toric Code model}

The 3D toric code model \cite{hamma2005string} is defined on a cubic lattice, with one qubit on each link. The Hamiltonian is of exactly the same form as the one of the 2D toric code model:
\[
H=-\sum_s A_s -\sum_{p} B_p;
\]
\[
A_s=\prod_{r\in s}X_r;\qquad  B_{p}=\prod_{r\in p} Z_r.
\]
Here a star $s$ includes the 6 links around a vertex, and a plaquette $p$ is a square consistent of 4 links.

\begin{figure}[h]
	\centering
	\includegraphics[scale=0.340]{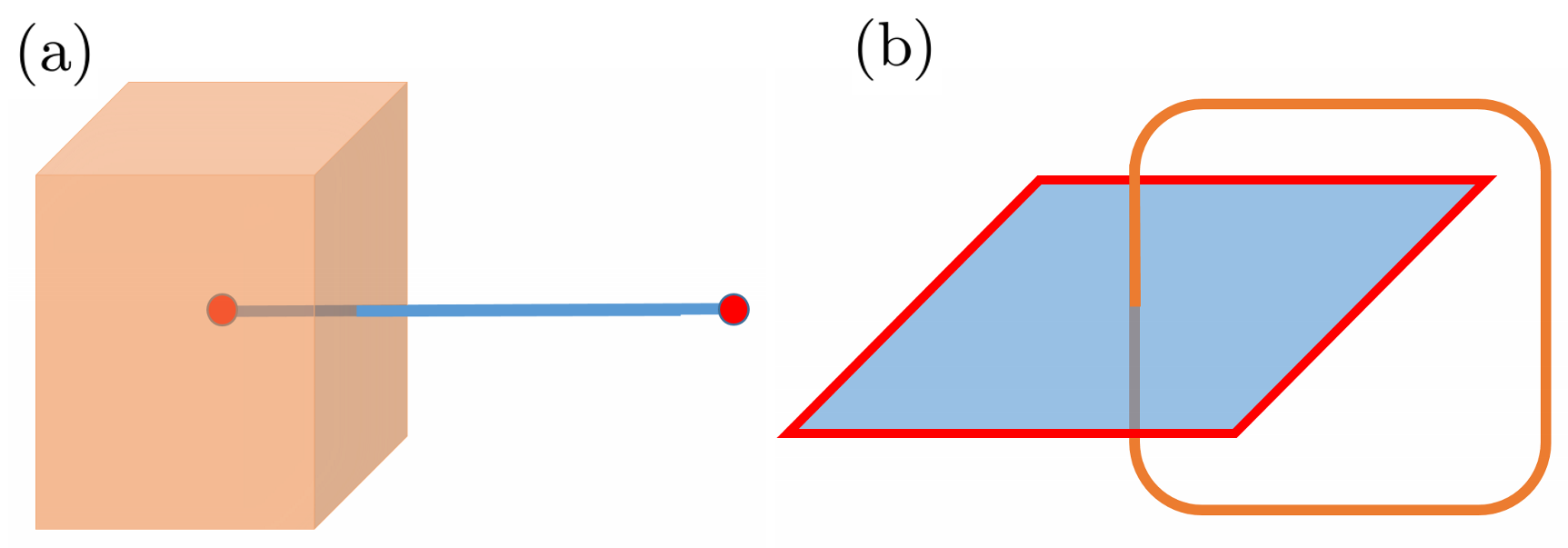}
	\caption{Topological excitations in the 3D toric code model and their detection using condensates. }\label{3D Toric Code}
\end{figure}

The ground state of the 3D toric code model condenses one type of closed string and one type of closed membrane. There is one type of open string operator that creates point-like topological excitations at the endpoints and one type of open membrane operator which creates loop-like topological excitations at the edge of the membrane, see Fig.\ref{3D Toric Code}.

\subsubsection{Subsystem types for 3D models}
\begin{figure}[h]
	\centering
	\includegraphics[scale=0.320]{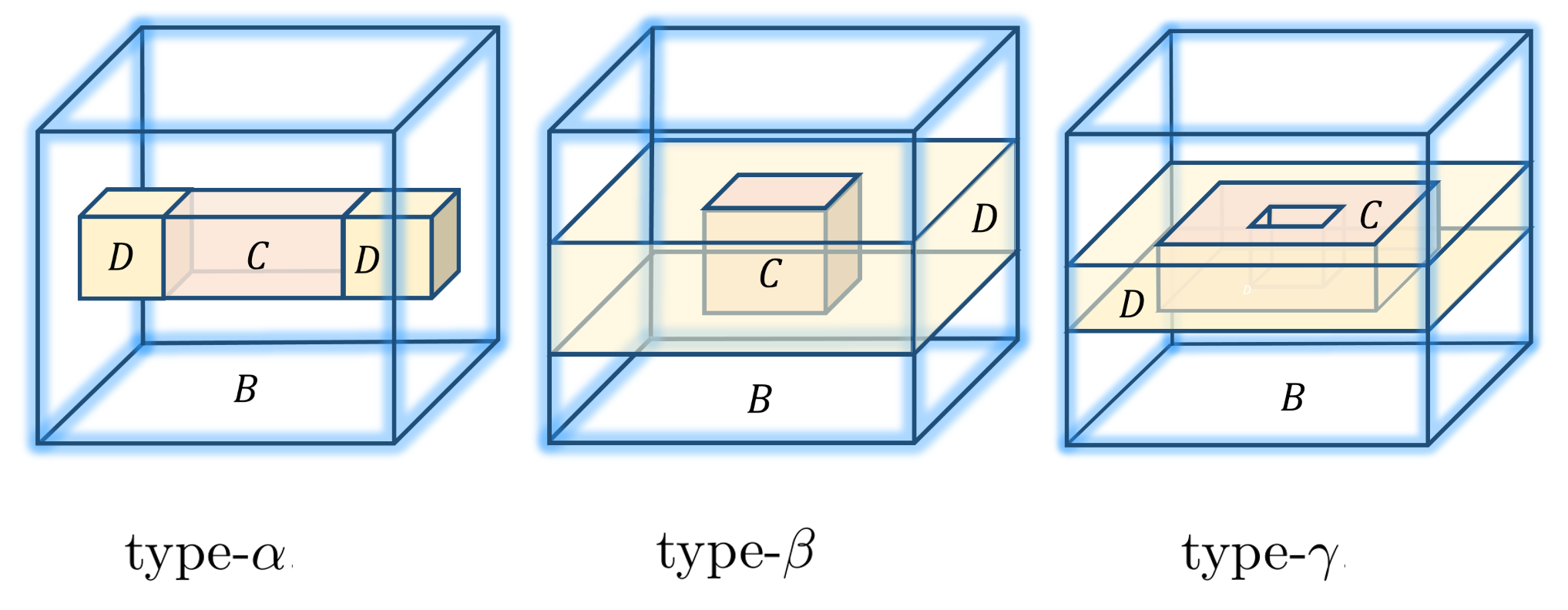}
	\caption{Three topologically distinct choices of subsystems, e.g. type-$\alpha$, type-$\beta$ and type-$\gamma$. We make subsystem $B$ transparent in order to see $C$ and $D$ more clearly. It is understood that $B\cup CD$ is the box with blue edges and $A$ is the complement of the box. We also apply this convention to pictures below.}\label{3 types}
\end{figure}

The 3D toric code model is the first 3D model we discuss, and it is a good place to introduce subsystem types for 3D models which  will be discussed for all 3D models. We  focus on the following three topologically distinct  subsystem types, i.e. the type-$\alpha$,$\beta$,$\gamma$ shown in Fig.\ref{3 types}, although other choices are possible. We will use the notation $S^{(\alpha)}_{nonlocal}$, $S^{(\beta)}_{nonlocal}$, $S^{(\gamma)}_{nonlocal}$ to distinguish the nonlocal entanglement entropy for the three topological types.

Type-$\alpha$ has  $D$, which consists of two disconnected boxes, while $C D$ is connected, and it can be used to detect open string-like $U_i$,  which is attached to the two boxes of $D$.
Type-$\beta$ has $D$ of the topology of a solid torus (and therefore $D$ is not simply connected), while $C D$ is simply connected.
It can be used to detect open membrane-like $U_i$ supported on $C D$ which create excitations in $D$ as noncontractible loops.
Type-$\gamma$ has $C$ and $CD$ of the same topology, e.g. the topology of a solid torus, and $B$ is simply connected.

Type-$\alpha$ and type-$\beta$ have already been implied in paper by Kim and Brown \cite{2015PhRvB..92k5139K}, in which, similar subsystems types are used to study different types of boundaries of 3D models.

For type-$\gamma$, $S_{nonlocal}^{(\gamma)}=0$ for models satisfying the assumptions in \cite{PhysRevB.84.195120}, e.g. the entanglement entropy of a general subsystem $\Omega$ (which has  large size compared to correlation length) can be decomposed into local  plus topological parts:
\[S_{\Omega}= S_{\Omega,local} +S_{\Omega,topological}\quad \Rightarrow \quad S_{nonlocal}^{(\gamma)}=0.\]
Therefore, a model with $S_{nonlocal}^{(\gamma)}>0$  is a model beyond the description of  \cite{PhysRevB.84.195120}. Fractal models do have $S^{(\gamma)}_{nonlocal}>0$ for some choices of the subsystems and the value can be extensive.  Furthermore, the contribution of $S^{(\alpha)}_{nonlocal}$ ($S^{(\beta)}_{nonlocal}$) is not necessarily from open string-like (open membrane-like) $U$.

\subsubsection{The 3D Toric Code model has saturated lower bounds for each subsystem type}
Let us go back to the 3D toric code model.

For type-$\alpha$,  we find $M=1$, $n_1=2$ where the operator $U_1$ is an open string operator which creates a point-like topological excitation in each box of $D$ and $W_1$ is a closed membrane operator. Therefore $N=2$ and $S_{nonlocal}^{(\alpha)}\ge \ln 2$.

For type-$\beta$, we find $M=1$, $n_1=2$ where the operator $U_1$ is an open membrane operator which creates a loop-like topological excitation at the edge of the membrane (the loop could not continuously shrink within $D$ into a point), and $W_1$ is a closed string operator. Therefore $N=2$ and $S_{nonlocal}^{(\beta)} \ge \ln 2$.

For type-$\gamma$, operators supported on $CD$, which create excitations in $D$, could always be deformed into $D$. This is because, for 3D toric code, the operators that create topological excitations  can be topologically deformed keeping the excitations fixed. Therefore we obtain a lower bound  $S_{nonlocal}^{(\gamma)}\ge 0$.

Comparing with the known entanglement properties \cite{PhysRevB.84.195120,2015PhRvB..92k5139K} of the 3D toric code model, our lower bounds are identical to the exact results:
\[ S^{(\alpha)}_{nonlocal}=S^{(\beta)}_{nonlocal}=\ln 2;\qquad S^{(\gamma)}_{nonlocal}=0.
\]

\subsection{The X-Cube Model}\label{X-Cube}
The X-cube model is a 3D exactly solved stabilizer model, and it is an example of type \Rom{1} fracton phase \cite{PhysRevB.94.235157}.
The model is defined on a cubic lattice with one qubit on each link. The Hamiltonian is
\begin{equation}
H=-\sum_{c} A_c -\sum_s (B^{(xy)}_s + B^{(yz)}_s +B^{(zx)}_s),
\end{equation}
where $A_c$ is a product of $Z_r$ on a cube (which includes 12 links), and $B^{(xy)}_s$, $B^{(yz)}_s$, and $B^{(zx)}_s$ are products of $X_r$  of 4 links around a vertex which are parallel to the $xy$-plane, $yz$-plane, $zx$-plane respectively.

\begin{figure}[h]
	\centering
	\includegraphics[scale=0.430]{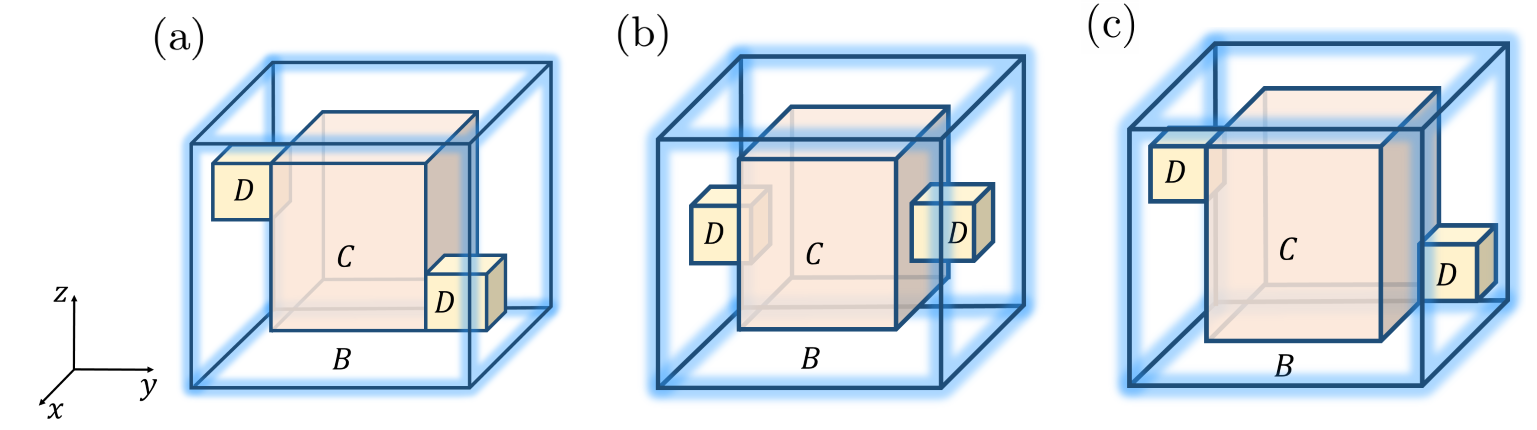}
	\caption{Type-$\alpha$ subsystem choices: The two boxes of $D$ are of the same size $l_D\times l_D \times l_D$, and they are separated by a displacement vector $\vec{d}=(d_x,d_y,d_z)$. $\partial D \cap \partial A$ contains two 2D pieces parallel to the $xz$-plane. (a) $d_x=0$, $\vert d_y\vert,\vert d_z\vert >l_D$. (b) $d_x=d_z=0$, $\vert d_y\vert >l_D$. (c) $\vert d_x\vert ,\vert d_y\vert , \vert d_z\vert >l_D$.}\label{string2}
\end{figure}
Here we focus on  type-$\alpha$ and consider the  geometry dependence of $S_{nonlocal}$, see Fig.\ref{string2}.
We find the following lower bounds:

Fig.\ref{string2}a, $S^{(\alpha)}_{nonlocal}\ge (2l_D +O(1))\ln 2$;

Fig.\ref{string2}b, $S^{(\alpha)}_{nonlocal}\ge (4l_D +O(1))\ln 2$;

Fig.\ref{string2}c, $S^{(\alpha)}_{nonlocal}\ge 0$.\\
Where $O(1)$ denotes  order one contributions which dependent on the detailed shapes of the subsystems, which is not crucial for our discussion.

\begin{figure}[h]
	\centering
	\includegraphics[scale=0.400]{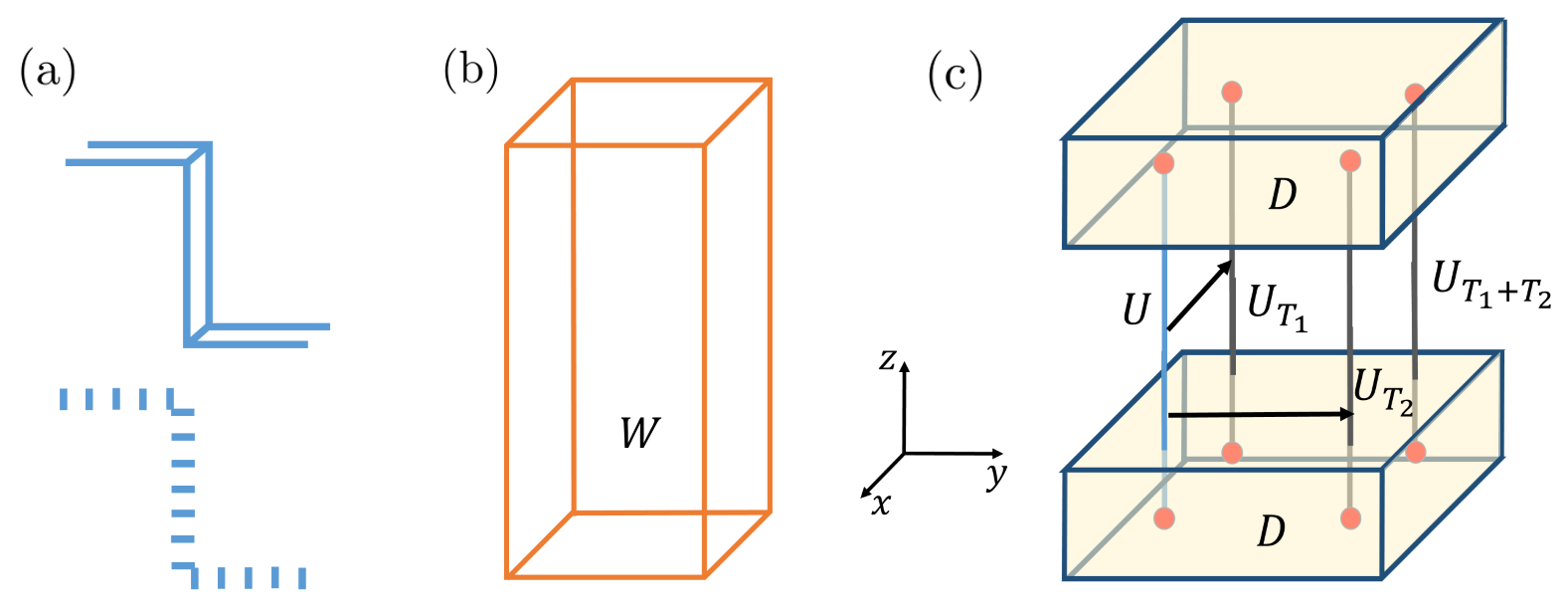}
	\caption{About $U$ and $W$ in the X-cube model: (a) The two types of $U$ and their translations in $(1,0,0)$ direction contribute to $S_{nonlocal}$ for the configuration in Fig.\ref{string2}a. (b) A product of $A_c$ gives you a $W$ that is a product of $Z_r$ on the 1D edges of a cuboid. (c) $U$ is a product of $Z_r$ on a line parallel to $z$-axis and it creates point-like excitations in $D$. $U_{T_1}$, $U_{T_2}$, and  $U_{T_1+T_2}$ are translations of $U$ by vectors $\vec{T}_1=(-a,0,0)$, $\vec{T}_2=(0,b,0)$, and $\vec{T}_1+\vec{T}_2=(-a,b,0)$, and $a$ and $b$ are positive integers in unit of lattice spacing. }\label{X Cube String}
\end{figure}

The types of $U_i$ that contribute to the $S^{(\alpha)}_{nonlocal}$ in Fig.\ref{string2}a are illustrated in Fig.\ref{X Cube String}a, each $U_i$ stretches out in directions parallel to the $yz$-plane and creates a pair of ``dimension-2 anyons." The translations of the operators $U_i$  in Fig.\ref{X Cube String}a in $(1,0,0)$ direction give you distinct operators (while translations in $(0,1,0)$ or $(0,0,1)$ directions do not give you distinct operators). This gives $M=2l_D +O(1)$.

Translations can produce distinct operators, this indicates a breakdown of topological deformation: in the X-cube model, $U_i$  is deformable but not topologically deformable. Another nice example of the breakdown of topological deformation is shown in Fig.\ref{X Cube String}c, in which $U$ and its translations $U_{T_1}$, $U_{T_2}$, and $U_{T_1+T_2}$ are distinct; nevertheless, by thinking of the condensate in Fig.\ref{X Cube String}b, one can show $U\sim U_{T_1}\cdot U_{T_2}\cdot U_{T_1 +T_2}$.

The result $S^{(\alpha)}_{nonlocal}\ge (4l_D +O(1))\ln 2 $ for Fig.\ref{string2}b can be understood by thinking of contributions from operators parallel to the $yz$-plane and the $xy$-plane, which gives $M=4l_D +O(1)$. The result  $S^{(\alpha)}_{nonlocal}\ge 0$ in Fig.\ref{string2}c comes from the fact that the types of $U_i$ discussed above could not connect the two boxes of $D$ separated by a displacement vector $\vec{d}=(d_x,d_y,d_z)$ with $\vert d_x\vert, \vert d_y\vert, \vert d_z\vert >l_D$ and the inability to find $U_i$ gives $M=0$.

These results for $S^{(\alpha)}_{nonlocal}$ may also be calculated using the method in \cite{PhysRevA.71.022315} and an independent estimation agrees with our lower bounds up to $O(1)$ contributions.

The lower bounds of $S^{(\alpha)}_{nonlocal}$ for all the cases in Fig.\ref{string2} depend on the length scale $l_D$ and the displacement vector $\vec{d}$ but are not sensitive to other details. This is due to the fact that we have chosen ``big enough" $A,B,C$, so that they do not block any $U_i$. If we consider another extreme, say the subsystem $C$ has a very narrow neck, then $S^{(\alpha)}_{nonlocal}$ will be sensitive to the geometry of the neck which determines how many $U_i$ could pass through.

One may also apply the same idea to subsystems of  type-$\beta$ and type-$\gamma$ and find extensive values of $S^{(\beta)}_{nonlocal}$ and $S^{(\gamma)}_{nonlocal}$ for certain choices of subsystems.

\subsection{Fractal spin liquids}\label{fractal model}
Fractal spin liquids \cite{PhysRevB.88.125122} is a generalization of Haah's code \cite{PhysRevA.83.042330}. A common feature of fractal spin liquid models is the existence of fractal condensates. Fractal structures have discrete scale symmetries, and this results in a more complicated dependence of the ground state degeneracy on the system size \cite{Haah2013,PhysRevB.88.125122} compared to the type  \Rom{1} fracton models.

Some of the fractal models possess ``hybrid" condensates $W_i$ having both 1D parts and fractal parts, and the truncations of the condensates give you $U_i$ which can be either a string-like operator or a fractal operator. Note that, they do not fit into the definition of type \Rom{1} due to the existence of fractal operators. On the other hand, they do not fit into type \Rom{2} because the excitations created by the string-like operator are mobile excitations. Some fractal models have only fractal condensates, and no string-like $U_i$ exists. These models are type \Rom{2} fracton models.

Discussed in the following are  ways to detect string-like $U_i$ and fractal $U_i$ using condensates. Then, $S_{nonlocal}$ is shown to be extensive for certain choices of subsystem geometry.

\subsubsection{The Sierpinski Prism Model}
\begin{figure}[h]
	\centering
	\includegraphics[scale=0.380]{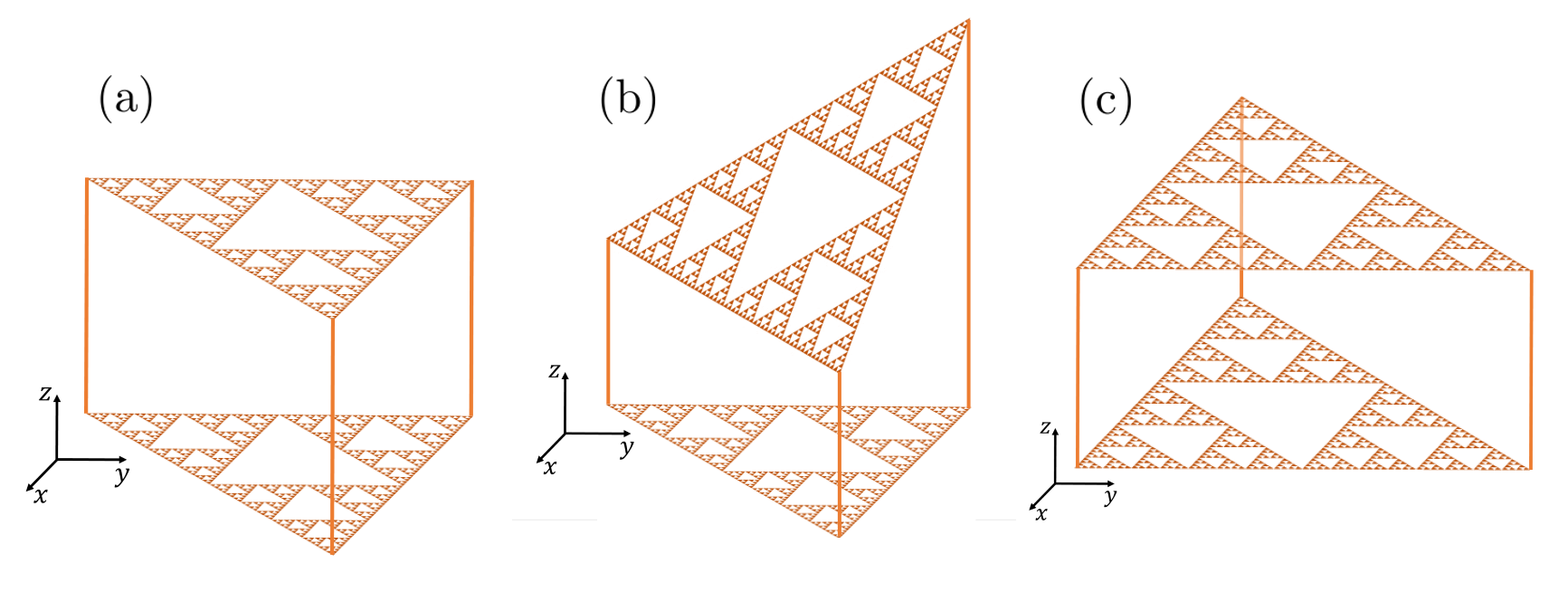}
	\caption{Examples of condensates in the Sierpinski prism model: (a) The fractal structures (here are Sierpinski triangles) contained in the upper and lower surfaces parallel to the $xy$-plane are products of $Z_r^B$. The three strings parallel to the $z$-axis are products of $Z_r^A$. (b) A condensate that can be thought of as a ``deformed version" of Fig.\ref{Fractal Condensates}a, and now the upper surface is a Sierpinski triangle perpendicular to (1,-1,1) direction and it is a product of both $Z^A_r$ and $Z_r^B$, its lower surface is the same as that in Fig.\ref{Fractal Condensates}a, i.e. a product of $Z_{r}^B$. (c) A condensate different from the former two, its upper and lower surfaces parallel to the $xy$-plane are products of $X^A_r$ and the strings parallel to $z$-axis are products of $X^B_r$.}\label{Fractal Condensates}
\end{figure}

As an example, we consider the  model (d) in Yoshida's paper \cite{PhysRevB.88.125122}. Let us call this model the \emph{Sierpinski prism model}, named after the shape of the condensate in Fig.\ref{Fractal Condensates}a, which looks like a prism with three legs decorated with Sierpinski triangles. This model lives on a 3D cubic lattice with two qubits ($A$ and $B$) on each site. The Hamiltonian can be written as
\begin{equation}
H=-\sum_{i,j,k} h^Z_{(i,j,k)}-\sum_{i,j,k} h^X_{(i,j,k)},
\end{equation}
where $i,j,k$ are integers labeling the sites on cubic lattice, and the Hamiltonian involves all the translations of the operator $h^Z_{(0,0,0)}$ and $h^X_{(0,0,0)}$. Explicitly, in terms of Pauli operators acting on each $A$ and $B$ qubit on different sites, we have
\begin{eqnarray}
h^Z_{(0,0,0)}&=&Z^A_{r=(0,0,0)}Z^A_{r=(0,1,0)}Z^A_{r=(1,1,0)}\nonumber
\\&&\times Z^B_{r=(0,0,0)}Z^B_{r=(0,0,1)}\nonumber\\
h^X_{(0,0,0)}&=&  X^A_{r=(0,0,0)}X^A_{r=(0,0,-1)}\nonumber
\\&&\times X^B_{r=(0,0,0)}X^B_{r=(0,-1,0)}X^B_{r=(-1,-1,0)}. \nonumber
\end{eqnarray}
It is easy to check that all terms in the Hamiltonian commute and $[h_{(i,j,k)}^Z]^2=[h^X_{(i,j,k)}]^2=1$.

Using Yoshida's notation:
\[
h^Z_{(0,0,0)}=  Z \left(\begin{array}{c}
1+y+xy\\ 1+z
\end{array}\right) ;\quad
h^X_{(0,0,0)}= X \left(\begin{array}{c}
1+\bar{z}\\ 1+\bar{y}+\bar{x}\bar{y}
\end{array}\right).
\]
Where $\bar{x}\equiv x^{-1}$, $\bar{y}\equiv y^{-1}$ and $\bar{z}\equiv z^{-1}$.
In terms of polynomials $f(x)$, $g(x)$ with coefficients over $\mathbf{F}_2$, i.e.  the coefficients can take 0 or 1:
\[
h^Z_{(0,0,0)}=  Z \left(\begin{array}{c}
1+f(x)y\\ 1+g(x)z
\end{array}\right) ;\quad
h^X_{(0,0,0)}= X \left(\begin{array}{c}
1+\bar{g}(x)\bar{z}\\ 1+\bar{f}(x)\bar{y}
\end{array}\right).
\]
Here $f(x)=1+x$ and $g(x)=1$ for the Sierpinski prism model. The polynomials with $\mathbf{F}_2$ coefficients indicate the locations and the numbers of Pauli $Z$ or $X$ operators in the product; the upper row is for  $A$ qubits and the lower row is for $B$ qubits.

Choosing other polynomials $f(x)$, $g(x)$ or changing  $\mathbf{F}_2$ into $\mathbf{F}_p$ ($p>2$ prime number) will generally give you other fractal models.

The Sierpinski prism model possesses hybrid condensates which consist of 1D parts and fractal parts, see Fig.\ref{Fractal Condensates}. The condensates in Fig.\ref{Fractal Condensates}a and Fig.\ref{Fractal Condensates}b can be constructed as a product of $h^Z_{(i,j,k)}$ and the condensate in Fig.\ref{Fractal Condensates}c can be constructed as a product of $h^X_{(i,j,k)}$. As is suggested by the discrete scaling symmetry of fractal structure and the continuous scaling symmetry of a 1D line: the upper and lower surfaces can be separated by an arbitrary distance in $z$-direction (without changing the size of upper/lower surfaces), and under a rescaling $l\to 2^m l$ the condensates look similar. Under other rescaling factors, the condensates look different but they could be constructed using a product of condensates that looks similar to the ones in Fig.\ref{Fractal Condensates}.

While this model does not have any logical qubits under periodic boundary conditions on an $L_x\times L_y\times L_z$ lattice, i.e. $x^{L_x}=y^{L_y}=z^{L_z}=1$, it does have logical qubits under some ``twisted" boundary conditions (say $x^{L_x}=1$, $y^{L_y}=x$, $z^{L_z}=1$ with $L_x=2^m+1$, $L_y=2^m$, $\forall\, L_z$ and integer $m$), or under open boundary conditions.

Despite the fact that the Sierpinski prism model is one of the simplest fractal models, it nicely illustrates all the important ingredients needed in order to understand how our method works in fractal models. To be specific, it illustrates the following three types of detections:\\
1) The detection of a string-like $U$ using a fractal  $W$.\\
2) The detection of a fractal $U$ using a string-like  $W$.\\
3) The detection of a fractal $U$ using a fractal $W$.
\subsubsection{The detection of sting-like $U$  by fractal $W$}

\begin{figure}[h]
	\centering
	\includegraphics[scale=0.360]{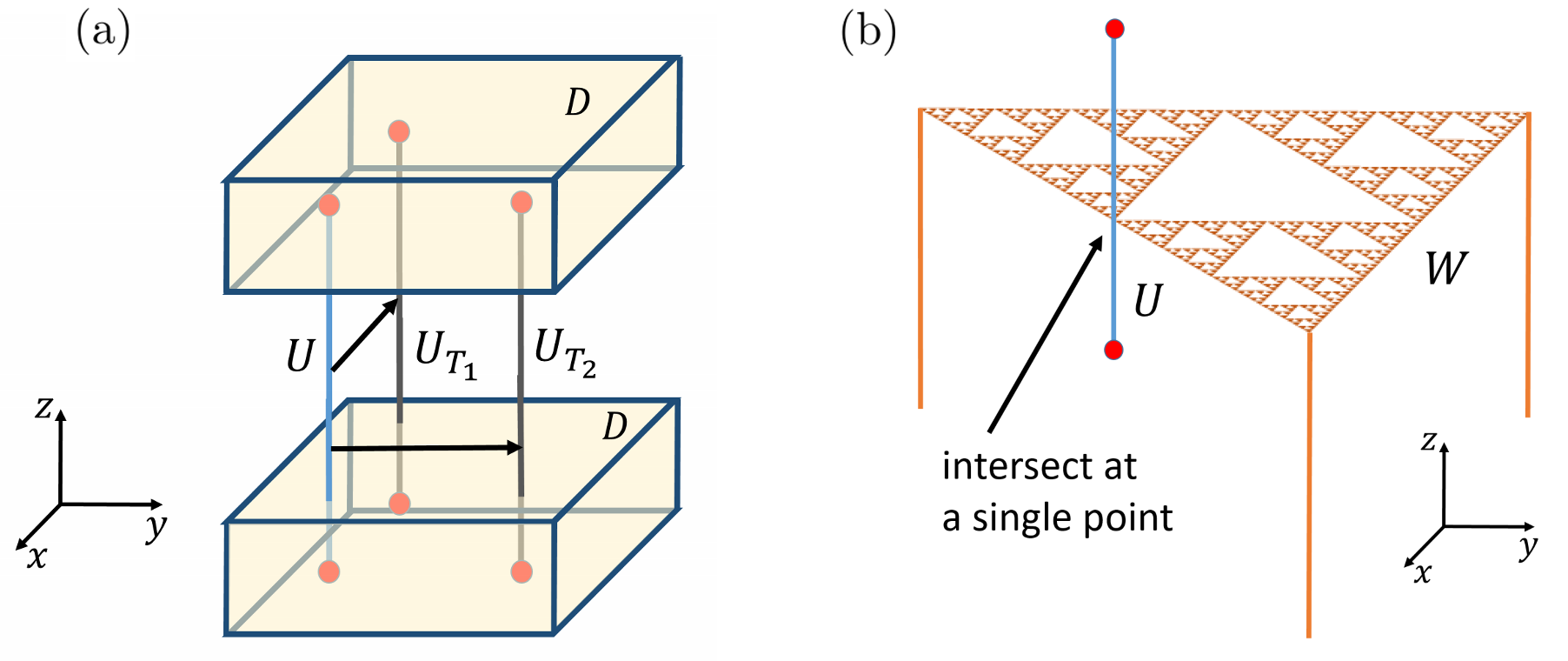}
	\caption{A string-like $U$ which can be thought of as a truncation of the condensate in Fig.\ref{Fractal Condensates}c. It creates point-like excitations at its endpoints. (a) Translating $U$ by vectors $\vec{T}_1$, $\vec{T}_2$, which naturally appear in the fractal structure, to get $U_{T_1}$ and $U_{T_2}$, where $\vec{T}_1=(-2^m,0,0)$ and $\vec{T}_2=(0,2^m,0)$ with an integer $m$. (b) The detection of $U$ using the fractal part of a condensate of the same type as Fig.\ref{Fractal Condensates}a.}\label{Fractal String}
\end{figure}
As is shown in Fig.\ref{Fractal Condensates}, we have condensates with string-like parts and fractal parts. String-like $U_i$ can be obtained from a truncation of the condensates. In Fig.\ref{Fractal String}a, translations of a string-like $U$, e.g. $U_{T_1}$ and $U_{T_2}$ (with $\vec{T}_1=(-2^m,0,0)$, $\vec{T}_2=(0,2^m,0)$ and an integer $m$)  are distinct from $U$, but $U\sim U_{T_1}\cdot U_{T_2}$. It is different from what happens in conventional topological orders but similar to what happens in type \Rom{1} fracton models, see Fig.\ref{X Cube String}.

The distinctness of a string-like $U$ and its translations indicates a lower bound of $S_{nonlocal}$ extensive in the subsystem size, and indeed we can use the fractal part of $W_i$ to detect different string-like $U_i$, see Fig.\ref{Fractal String}b, and get an extensive lower bound.

\subsubsection{The detection of fractal $U$ by string-like $W$ }
\begin{figure}[h]
	\centering
	\includegraphics[scale=0.420]{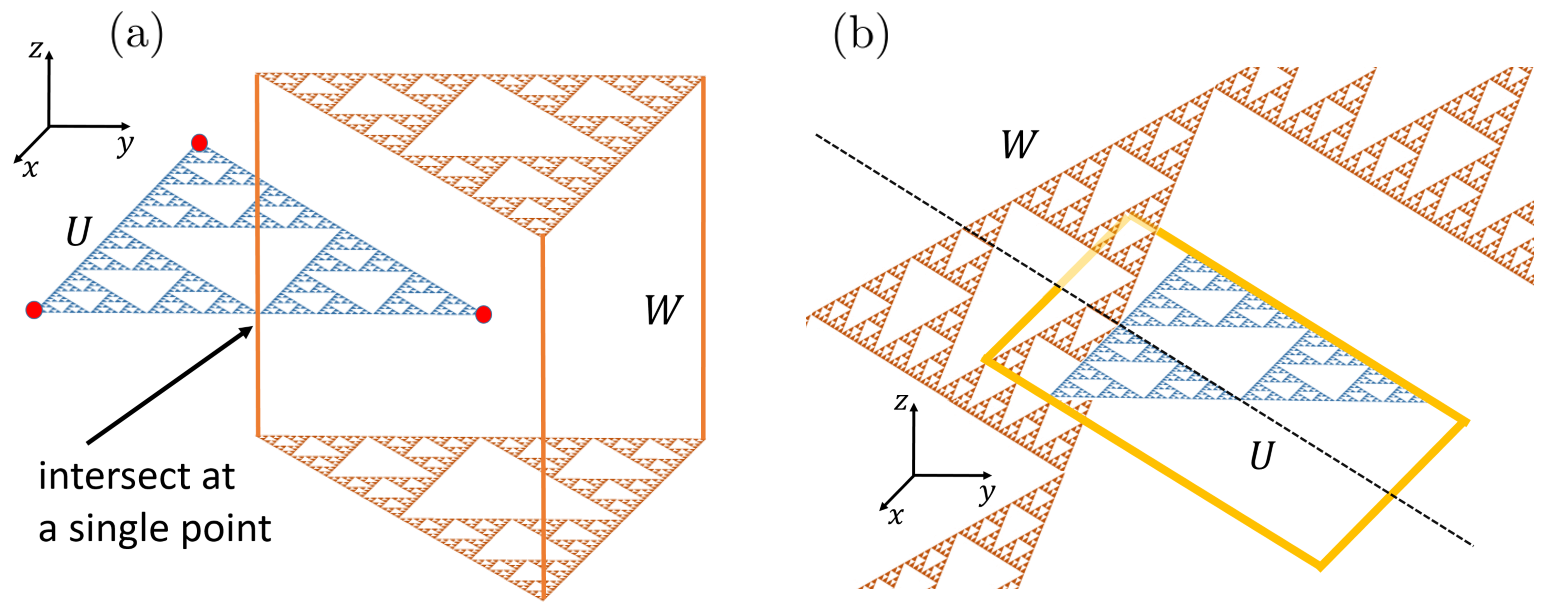}
	\caption{The fractal $U$ comes from a truncation of the condensate in  Fig.\ref{Fractal Condensates}c, and it creates point-like excitations at the three vertices of the Sierpinski triangle. (a) Detecting $U$ using the string-like part of $W$ of the type in Fig.\ref{Fractal Condensates}a. (b) Detecting $U$ using the fractal part of a $W$ similar to the upper surface of Fig.\ref{Fractal Condensates}b. The fractal structure of $U$ and $W$ lie in different planes and the dashed line is the intersection of the two planes. Shown in yellow color is a choice of subsystem $D$ for  type-$\beta$ or type-$\gamma$.}\label{Fractal Fractal}
\end{figure}

Very similarly, fractal $U$ can be detected by string-like parts of $W$, see Fig.\ref{Fractal Fractal}a. It is clear that a translation of a fractal $U$ will be a distinct operator if it anticommutes with a different $W$. Therefore, by suitably choosing the geometrical shapes of subsystems $A, B, C, D$, it is possible to get an extensive lower bound of $S_{nonlocal}^{(\alpha)}$. $S_{nonlocal}^{(\beta)}$ and $S_{nonlocal}^{(\gamma)}$.

\subsubsection{The detection of fractal $U$ by fractal $W$}
Another way to detect fractal $U$ is to use a fractal part of a condensate $W$, see Fig.\ref{Fractal Fractal}b, and it is the only way to detect $U$ for those fractal models without string operators i.e., when the condensates contain only a fractal structure without string-like parts (i.e.,  type \Rom{2}). Therefore, it is important to understand this case.

The key features, which can be observed in Fig.\ref{Fractal Fractal}b are the following.

1) The fractal condensates are  supported on 2D surfaces with ``holes" of different (discrete) length scales, in other words it is  less than 2D.

2) Fractal $U$ and fractal (part of) $W$ lie in distinct intersecting surfaces, and it is possible to make the operators intersecting at a point. Certain translation of $U$ has a non-overlapping support with $W$ and therefore it commutes with $W$. This implies that translations of $U$ can give you distinct operators.

After some thought, one finds it is  possible to get an extensive lower bound of $S_{nonlocal}^{(\alpha)}$, $S_{nonlocal}^{(\beta)}$ and $S_{nonlocal}^{(\gamma)}$ by suitably choosing the geometrical shapes of the subsystems $A, B, C, D$. A choice of $D$ for type-$\beta$ or type-$\gamma$ is shown in  yellow color in Fig.\ref{Fractal Fractal}b.
\subsubsection{Further comments}

1) One may consider other subsystem types, for example: type-$\delta$ with $D$ consists of three disconnected boxes, and  suitably chosen $A,B,C$. When putting the three boxes on the positions of the three excitations of a $U$ in Fig.\ref{Fractal Fractal},  $S^{(\delta)}_{nonlocal}>0$.

2) When rescaling the subsystem sizes according to the discrete scale symmetries (for the Sierpinski prism model, it is $L\rightarrow 2^m L$), the change of $S_{nonlocal}$ could be investigated using entanglement renormalization group transformation \cite{haah2014bifurcation}. For a rescaling by a factor which is not in the discrete scaling group, $S_{nonlocal}$ may change in more complicated way.

3) It is possible to have a fractal model with a unique ground state on a $T^3$, in which case, there still exists nonzero $S_{nonlocal}$. This indicates that $S_{nonlocal}$ is in some sense more universal than the ground state degeneracy.

4) For models with only fractal condensates (i.e. type \Rom{2}), although it is possible to find $S_{nonlocal}^{(\alpha)}> 0$ for some choices of $A, B, C, D$, we get a lower bound $0$ when, for example, the two boxes of $D$ of length $l_D\times l_D\times l_D$ are separated by a displacement vector $\vec{d}$ satisfying $\vert\vec{d}\vert>\lambda l_D$ with some constant $\lambda$ depending on the model. It might be a general exact result that  $S^{(\alpha)}_{nonlocal}=0$ when $\vert\vec{d}\vert>\lambda l_D$ no matter how you choose $A,B,C$ but we do not have a proof. The case for Haah's code is a conjecture by Kim \footnote{I. H. Kim, private communication. }. If the conjectured results are true, this may be used to make a clear distinction between type \Rom{1} and type \Rom{2} fracton models.

\section{Perturbations}\label{Sec.3}
The stability of quantities under local perturbations is an extremely important topic. If some property of an exactly solved model is totally changed when a tiny local perturbation is added,  this property could never be observed in real systems.

The ground state degeneracy of topological orders is robust (stable) to arbitrary local perturbations. It is known that the toric code model is stable under arbitrary local perturbations \cite{Kitaev20032}. The stability of ground state degeneracy is proved \cite{bravyi2010topological,bravyi2011short} for a very general class of models which satisfy assumptions TQO-1 and TQO-2. In the proof, Osborne's modification \cite{osborne2007simulating} of the quasi-adiabatic continuation \cite{hastings2005quasiadiabatic} is employed. This proof is applicable to both conventional and fracton topological orders.

We would like to understand the stability of $S_{nonlocal}$ under local perturbations. It turns out that the stability of $S_{nonlocal}$ is a trickier problem compared to the stability of the ground state degeneracy. The corresponding problem for the conventional topological orders, e.g. the stability of topological entanglement entropy is  not solved completely  without additional assumptions. It is known from Bravyi's counterexample (see \cite{zou2016spurious} for a published reference) that the arguments  provided in the original works \cite{PhysRevLett.96.110404,PhysRevLett.96.110405}  about the invariance of the topological entanglement entropy under perturbation are not complete. Kim obtained a bound of the change of topological entanglement entropy with a $1$st order perturbation \cite{PhysRevB.86.245116} assuming the conditionally independence of certain subsystems.

Here we study the stability of our lower bound of $S_{nonlocal}$ under a finite depth quantum circuit for simplicity, since it is known from the viewpoint of quasi-adiabatic evolution \cite{hastings2005quasiadiabatic} that local perturbations for gapped systems can be approximated by finite depth quantum circuit \cite{chen2010local,haah2016invariant}.

\begin{figure}[h]
	\centering
	\includegraphics[scale=0.4]{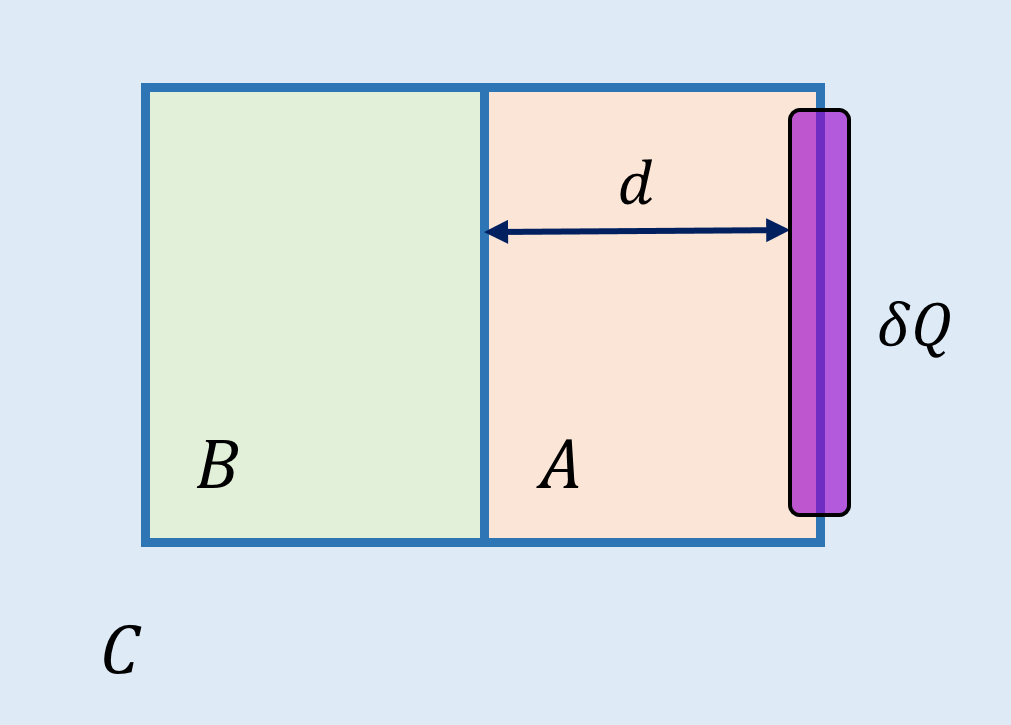}
	\caption{Subsystems $A$, $B$, $C$ and the unitary operator $\delta Q$ (in purple color) which appears in assumption $\mathbf{S}$. $\delta Q$ is separated from $B$ by a distance $d$.}\label{S condition}
\end{figure}

\emph{Assumption} $\mathbf{S}$: For subsystems $A$, $B$, $C$ as is shown in Fig.\ref{S condition}. $\delta Q$ is a unitary operator which has support intersecting with $A$ and $C$. $B$ is separated from $\delta Q$ by a distance $d$. $\sigma$ is a density matrix of a state with correlation length $\xi$ and replica correlation length \cite{zou2016spurious} $\xi_\alpha$ and $\sigma'\equiv \delta Q\sigma \delta Q^\dagger$. And $\sigma$ is a density matrix such that
\begin{equation}
\big(S_{AB}-S_A \big)\vert_{\sigma'}\simeq \big(S_{AB}-S_A \big)\vert_{\sigma}, \label{S equation}
\end{equation}
where ``$\simeq$" means there is a correction that is negligible when $d$ is large compared to $\xi$ and $\xi_{\alpha}$.

The assumption $\mathbf{S}$ should be understood as an assumption about the density matrix $\sigma$.
It is trivial to check that ``$\simeq$" can be replaced by ``$=$" when $\sigma_{AB}=\sigma_A\otimes\sigma_B$, and it may seem intuitive that the difference between the left-hand side and the right-hand side of Eq.(\ref{S equation}) should decay as $e^{-d/\xi}$.
Nevertheless, the original suggestion \cite{PhysRevLett.96.110404} that $\mathbf{S}$ is true for $\xi \ll d$ is violated in Bravyi's counterexample. It is observed in  \cite{zou2016spurious} that this is due to the fact that the replica correlation length $\xi_\alpha$ is infinity for the cluster state in Bravyi's counterexample. When the cluster state is deformed, $\xi_\alpha$ becomes finite. For generic local perturbations without symmetry requirement, it is fine-tuned to have $\xi_{\alpha}=+\infty$ but $\xi_\alpha$ can be arbitrarily large compared to $\xi$.
Judging from a recent conjecture \cite{zou2016spurious} , $\xi_\alpha\ll d$ may be the condition required for $\mathbf{S}$ to be true.

In the following, we discuss the stability of our lower bound of $S_{nonlocal}$ under a depth-$R$ quantum circuit $Q$ which creates a perturbed ground state $\bar{\rho}$ satisfying $\mathbf{S}$. We take $R\sim \xi$, and assume $\xi$ and $\xi_\alpha$ much smaller than the length scales of the subsystems. This analysis does not cover all possible local perturbations (especially those with $\xi_{\alpha}\to +\infty$), but we believe it covers a large class of interesting local perturbations.

Let $Q$ be the depth-$R$ quantum circuit ($Q Q^\dagger=1$) which is responsible for the local perturbation. In other words, we assume the following objects in the perturbated model are related to the corresponding objects in the unperturbed model by

1) The new Hamiltonian: $\bar{H}=QHQ^\dagger$;

2) The new (dressed) operators: $\bar{U}_i=Q U_i Q^\dagger$, $\bar{U}_{i}^{def}=Q \bar{U}_{i}^{def}Q^\dagger$, and $\bar{W}_i = Q W_i Q^\dagger$;

3) The new ground state: $\vert\bar{\psi}\rangle = Q \vert\psi\rangle$;

4) The new density matrices: $\bar{\sigma}_I=Q \sigma_I Q^\dagger$ and $\bar{\rho}=Q\rho Q^\dagger$.

\begin{figure}[h]
	\centering
	\includegraphics[scale=0.240]{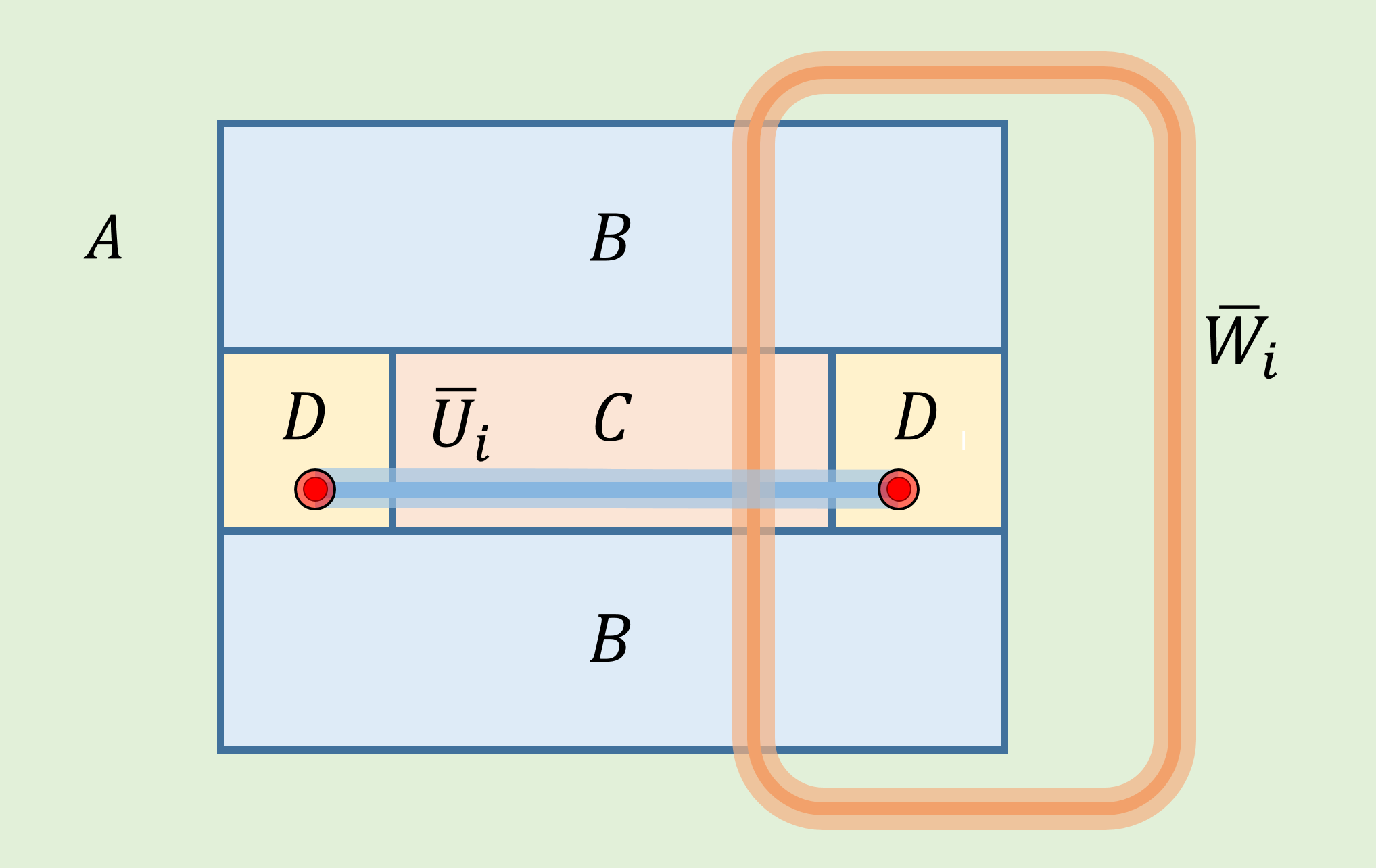}
	\caption{Dressed operators $\bar{W}_i$ and $\bar{U}_i$. Compare with the operators $W_i$ and $U_i$ shown in Fig.\ref{deform}, the width of the support of dressed operators can increase at most by $2R$ due to the depth-$R$ quantum circuit.}\label{dressed}
\end{figure}
The dressed operators typically have a ``fatter" support than the corresponding operators in the unperturbed stabilizer model, see Fig.(\ref{dressed}) for an illustration. It is possible that the support of some $\bar{U}_i$ will overlap with $AB$, the support of some $\bar{U}_{i}^{def}$ will overlap with $BC$ and the support of some  $W_i$ will overlap with $D$. For those operators, we need to throw them away, and supply with other operators if possible. We label the remaining operators using $\bar{i}$, $\bar{i}=1,\cdots \bar{M}$, where $\bar{M}\le M$, and we call the remaining density matrices $\bar{\sigma}_{\bar{I}}$, with $\bar{I}=1,\cdots, \bar{N}$, where $\bar{N}=\prod_{\bar{i}=1}^{\bar{m}}\bar{n}_{\bar{i}}$.

With $\mathbf{U}$-1, $\mathbf{U}$-2, $\mathbf{U}$-3, $\mathbf{W}$-1, $\mathbf{W}$-2  satisfied for the unperturbed system, we supply with $\mathbf{S}$ in order to complete a result about the stability of the lower bound. With these assumptions, one could verify the following results:\\
$1'$) $\mathbf{U}$-1, $\mathbf{U}$-3 $\Rightarrow$ $\bar{\sigma}_{\bar{I}\, AB}=\bar{\rho}_{AB}$ and $\bar{\sigma}_{\bar{I}\, BC}=\bar{\rho}_{BC}$;\\
$2'$) $\mathbf{W}$-1, $\mathbf{W}$-2 $\Rightarrow$ $\bar{\sigma}_{\bar{I}\, ABC}\cdot \bar{\sigma}_{\bar{J}\,ABC}=0$ for $\bar{I}\ne \bar{J}$;\\
$3'$) $\mathbf{U}$-1, $\mathbf{U}$-2, $\mathbf{U}$-$3$, $\mathbf{S}$ $\Rightarrow$ $S_{ABC}\vert_{\bar{\sigma}_{\bar{I}}}\simeq S_{ABC}\vert _{\bar{\rho}}$.

\begin{figure}[h]
	\centering
	\includegraphics[scale=0.26]{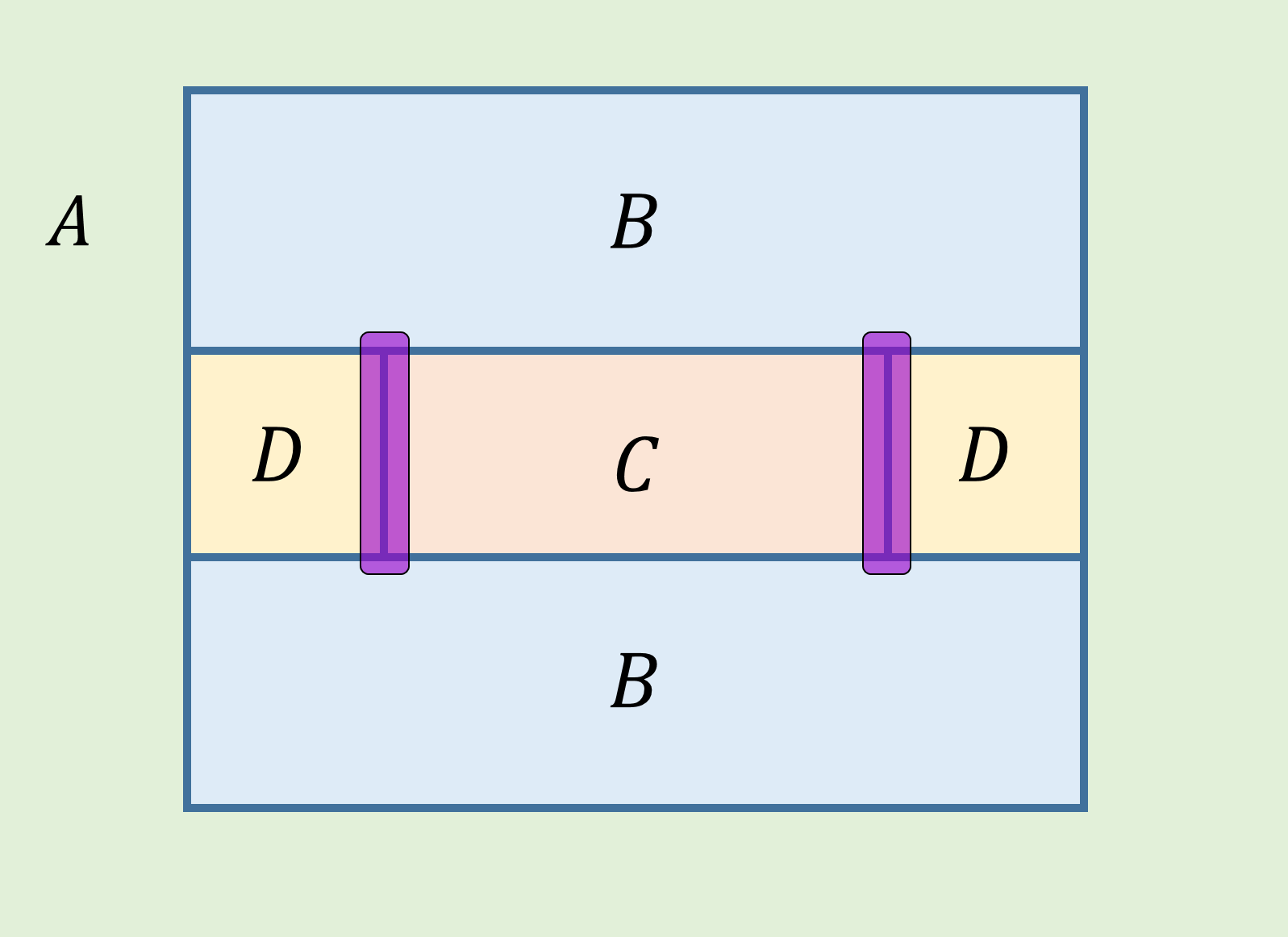}
	\caption{The support of the operator $\delta Q$ (in purple color) is within a distance $R$ around $\partial C \cap \partial D$.}\label{deltaQ}
\end{figure}

The derivation of $1'$) and $2'$) are parallel to what is done in Sec.\ref{The main result.}. The  derivation of $3'$) follows. There exists a unitary operator $\delta Q$ supported on a region within a distance $R$ around  $\partial C\cap \partial D$, See Fig.\ref{deltaQ}, such that  $\bar{\rho}=\delta Q \bar{\rho}' \delta Q^\dagger$, $\bar{\sigma}_{\bar{I}}=\delta Q \bar{\sigma}'_{\bar{I}}\delta Q^\dagger$ and $\bar{\sigma}'_{\bar{I}\,ABC}=\bar{V}_{\bar{I}}\bar{\rho}'_{ABC}\bar{V}^\dagger_{\bar{I}}$. $\bar{V}_{\bar{I}}$ is some unitary operator supported on $ABC$. It is always possible to find such $\delta Q$ and $\bar{V}_{\bar{I}}$ given that  $\mathbf{U}$-2 is satisfied for the unperturbed case.
Therefore, $S_{ABC}\vert_{\bar{\sigma}'_{\bar{I}}}=S_{ABC}\vert_{\bar{\rho}'}$. \\
Then, by applying  $\mathbf{U}$-$1$, $\mathbf{U}$-$3$, and $\mathbf{S}$ we find that
\begin{eqnarray}
\mathbf{U}\textrm{-}1, \mathbf{U}\textrm{-}3 &\Rightarrow& S_{BC}\vert_{\bar{\sigma}'_{\bar{I}}}=S_{BC}\vert_{\bar{\rho}'}\nonumber\\
&& S_{BC}\vert_{\bar{\sigma}_{\bar{I}}}=S_{BC}\vert_{\bar{\rho}}\,;\nonumber\\
\mathbf{S} &\Rightarrow&  (S_{ABC}-S_{BC})\vert_{\bar{\rho}}\simeq (S_{ABC}-S_{BC})\vert_{\bar{\rho}'}\nonumber\\ &&(S_{ABC}-S_{BC})\vert_{\bar{\sigma}_{\bar{I}}}\simeq (S_{ABC}-S_{BC})\vert_{\bar{\sigma}'_{\bar{I}}}\,.\nonumber
\end{eqnarray}
After simple algebra one arrives at the result $3'$) i.e.
$ S_{ABC}\vert_{\bar{\sigma}_{\bar{I}}}\simeq S_{ABC}\vert _{\bar{\rho}}$.

Assuming the error caused by ``$\simeq$" could be neglected, one could apply the same method as Sec.\ref{The main result.} to arrive at a lower bound
\begin{equation}
S_{nonlocal}\vert_{\bar{\rho}} \ge \ln \bar{N} =\sum_{\bar{i}=1}^{\bar{M}} \ln {\bar{n}}_{\bar{i}}.
\end{equation}

For models with topologically deformable $U_i$ operators, like the 2D, 3D toric code models, our lower bound is  invariant under perturbation. Because we can always move the excitations deep inside $D$, such that the distance from any excitation to the boundary $d_e\gg \xi\sim R$. For large subsystems,  we would have $M=M'$ and we do not lose any $U_i$, $U_{i}^{def}$ and $W_i$.

For models with $U_i$ not topologically deformable, e.g. the X-cube model and  fractal spin liquids. There is usually some excitation that could not be moved deep inside subsystem $D$. Therefore, after adding perturbations, we typically lose a few $U_i$ and $U_{i}^{def}$ , such that $\bar{M}<M$ and $\bar{N}<N$. But for large subsystems which possess extensive $S_{nonlocal}$ before perturbation is added, this modification is small comparing to the leading contribution.

To summarize, for local perturbations satisfying assumption $\mathbf{S}$, and subsystem sizes much larger than $\xi$ and $\xi_{\alpha}$ we expect:
\begin{equation}
S_{nonlocal}{(\textrm{perturbed})}\simeq
S_{nonlocal}{(\textrm{unperturbed})}-\mu\xi.
\end{equation}
For conventional topological orders $\mu=0$, and for fracton topological orders $\mu>0$ being a number depends on the model and subsystem geometry.

\section{Discussion and Outlook}\label{Sec.4}
In this paper, we have obtained a lower bound of the nonlocal entanglement entropy $S_{nonlocal}$ from assumptions about the topological excitations and the ground state condensates of Abelian topological orders and applied our method to several examples.
For conventional topological orders, e.g. the 2D toric code model and the 3D toric code model, our lower bounds are saturated and topologically invariant. Whenever the lower bound is saturated, we get an explicit construction of a conditionally independent density matrix $\sigma^\ast$.
For fracton topological orders \cite{PhysRevLett.94.040402,bravyi2011topological,PhysRevA.83.042330,PhysRevB.88.125122,PhysRevB.92.235136,PhysRevB.94.235157}, e.g. the X-cube model and the Sierpinski prism model, our lower bound  depends on the geometry of the subsystems and $S_{nonlocal}$ is extensive for certain subsystem choices.

This method observes an intimate relation between $S_{nonlocal}$ and the topological excitations and the ground state condensates, and it obtains a lower bound of $S_{nonlocal}$ without calculating the entanglement entropy of any subsystem. A nonzero lower bound of  $S_{nonlocal}$ is a result of the nonlocal nature of topological excitations, i.e. the fact that topological excitations could not be created alone by local operators. This nonlocal nature of topological excitations does not guarantee the operators which create the topological excitations to be topologically deformable and $S_{nonlocal}$ is not necessarily a topological invariance. Geometry-dependent $S_{nonlocal}$ is what appears in fracton models.  It is beyond an established paradigm, i.e., the topological entanglement entropy, and should be treated as its generalization. 
 The stability of the lower bound is discussed for local perturbations satisfying assumption $\mathbf{S}$, which should cover a large class of interesting local perturbations.

The different behaviors of $S_{nonlocal}$ may be used to distinguish fracton topological orders from conventional topological orders. Together with other  methods being developed so far \cite{PhysRevLett.111.080503,haah2014bifurcation}, our result provides  a better understanding of the entanglement properties of  fractal models. Furthermore, the lower bound suggests (but not prove) different behaviors of $S_{nonlocal}$  between type \Rom{1} and type \Rom{2} fracton models. These different behaviors may be proven or disproven by later works.

Some of the assumptions in our method do not apply to non-Abelian models, a variant of our lower bound of $S_{nonlocal}$ for non-Abelian models is presented in \cite{2018arXiv180101519S}. Also, it might be interesting to investigate possible implications of our method on relations among topological order, topological entanglement entropy and quantum black holes \cite{McGough2013,rasmussen2017gapless}.

\section*{Acknowledgement}
B.S. would like to thank Fuyan Lu for a comment on one assumption, Jeongwan Haah for providing a reference about Bravyi's counterexample, Isaac H. Kim for a discussion and sharing one conjecture, and Michael Levin for a discussion about the entanglement in non-Abelian phases. This work is supported by the startup funds at OSU and the National Science Foundation under Grant No. NSF DMR-1653769 (BS,YML).

\bibliography{ref}
\bibliographystyle{apsrev}

\end{document}